\documentclass{ws-ijmpe}
\usepackage[super,compress]{cite}
\usepackage{graphics}
\usepackage{graphicx,colordvi}
\usepackage{lscape,multirow}

\expandafter\ifx\csname package@font\endcsname\relax\else
\expandafter\expandafter
\expandafter\usepackage
\expandafter{\csname package@font\endcsname}
\fi
\DeclareRobustCommand\substyle{\name@idx{document substyle}}
\DeclareRobustCommand\classoption{\name@idx{document class option}}
\DeclareRobustCommand\classname{\name@idx{document class}}
\def\name@idx#1#2{{\ttfamily#2}
\index{#2\space#1=\string\ttt{#2}\space#1}\index{#1>#2=\string\ttt{#2}}}

\newcommand{\beq}[0]{\begin{equation}}
\newcommand{\eeq}[0]{\end{equation}}
\newcommand{\bea}[0]{\begin{eqnarray}}
\newcommand{\eea}[0]{\end{eqnarray}}
\newcommand{\minitab}[2][l]{\begin{tabular}{#1}#2\end{tabular}}
\usepackage{amssymb}

\def\d{\hbox{d}}
\def\be{\begin{equation}}
\def\ee{\end{equation}}
\def\bea{\begin{eqnarray}}
\def\eea{\end{eqnarray}}
\def\l{\label}

\def\hahat{\hat{H}}

\def\hahat0{\hat{H}_0}

\def\cos{\hbox{cos}}

\def\sinh{\hbox{sinh}}

\def\exp{\hbox{exp}}

\def\sinh{\hbox{sinh}}

\def\vareps{\varepsilon}

\def\siml{\hbox{\kern.1em \lower.6ex \hbox{$\sim$} \kern-1.12em
          \raise.6ex \hbox{$<$} \kern.1em }}
\def\simg{\hbox{\kern.1em \lower.6ex \hbox{$\sim$} \kern-1.12em
          \raise.6ex \hbox{$>$} \kern.1em }}

\def\siml{\hbox{\kern.1em \lower.6ex \hbox{$\sim$} \kern-1.12em
 \raise.6ex \hbox{$<$} \kern.1em}}
\def\simg{\hbox{\kern.1em \lower.6ex \hbox{$\sim$} \kern-1.12em
 \raise.6ex \hbox{$>$} \kern.1em}}

\newcommand{\beqar}{\begin{eqnarray}}
\newcommand{\eeqar}[1]{\label{#1} \end{eqnarray}}

\begin{document}
\markboth{A.G.~Magner, A.I.~Sanzhur, S.N.~Fedotkin, A.I.~Levon, S.~Shlomo}
{Shell-structure and asymmetry effects in level
  densities}

\catchline{}{}{}{}{}
\title{Shell-structure and asymmetry effects in level
  densities} 
\author{A.G. Magner, A.I. Sanzhur, S.N. Fedotkin, A.I. Levon}
\address{\it  Institute for Nuclear Research, Prospect Nauki 47, 03028 Kyiv,
  Ukraine}

\author{S. Shlomo}
\address{\it Cyclotron Institute, Texas A\&M University,
  College Station, Texas 77843, USA}

\maketitle

\begin{history}
\received{Day Month Year}
\revised{Day Month Year}
\end{history}

\begin{abstract}
  Level density $\rho(E,N,Z)$ is derived for a 
  nuclear system with
  a given energy $E$, neutron $N$, and proton $Z$ particle numbers,
  within the semiclassical
      extended Thomas-Fermi and periodic-orbit theory 
  beyond the Fermi-gas saddle-point method. We
  obtain $~~\rho \propto I_\nu(S)/S^\nu$,~~ where $I_\nu(S)$ is the
  modified Bessel function
  of the entropy $S$, 
  and $\nu$ is related to the number of integrals of motion,
  except for the energy $E$.
   For small shell structure contribution  one obtains
  within the 
  micro-macroscopic approximation (MMA)
the value of $\nu=2$
  for  $\rho(E,N,Z)$. 
   In the opposite case of much larger shell structure 
    contributions one finds a larger value of $\nu=3$.
      The MMA level density $\rho$ 
  reaches 
  the well-known 
    Fermi gas asymptote 
  for 
  large excitation energies,
  and the finite micro-canonical limit for
   low excitation energies. 
      Fitting the MMA $\rho(E,N,Z)$  to 
    experimental data on a long isotope chain
    for
  low excitation energies, due mainly to the shell effects,
  one obtains results for
   the
    inverse level
    density parameter $K$, 
    which differs 
    significantly from  that
     of
     neutron resonances. 
\end{abstract}

\ccode{PACS numbers: 21.10.-k, 21.10.Ma, 21.60.Cs, 24.10.Pa }

\section{Introduction}
\label{sec-introd}
The statistical level density is a fundamental tool for the description of 
many
properties of heavy nuclei
\cite{Be36,Er60,GC65,BM67,LLv5,Ig83,So90,Sh92,Ig98,AB00,AB03,AB15,EB09,AB16,ZS16,KZ16,HJ16,KS18,ZK18,ZH19,KZ20,KS20,FA21}.
Usually, the level density 
$\rho(E,N,Z)$, where $E$, $N$, and $Z$ 
are the energy, neutron,  and proton
numbers,
respectively,
is 
the inverse Laplace transformation of the partition function
$\mathcal{Z}(\beta,\alpha_n,\alpha_p)$.
Within the  grand
canonical ensemble, 
one can apply the standard Darwin-Fowler method
for the saddle-point 
integration
over all variables,
including 
$\beta$, which is related to the total energy $E$ \cite{Er60,BM67}.
This method assumes large
excitation energy $U$, so  that the temperature $T$ is
related to a  well-determined
saddle point
in the inverse Laplace integration variable
$\beta$ for a finite Fermi 
    system of large particle numbers. 
However,
many experimental data 
also 
exist for a low-lying part of the
excitation energy $U$,
where  such a saddle point does not exist; see, e.g., Ref.~\refcite{Le20}.
Therefore, the integral over the Lagrange multiplier
$\beta$ in the inverse Laplace
  transformation of the partition function
$\mathcal{Z}(\beta,\alpha_n,\alpha_p)$  
  should be carried out  more 
  accurately  beyond the standard saddle point method, 
  see
 Refs.~\refcite{KM79,NPA,PRC}. 
  However, for other variables
  related 
  to the  neutron $N$ and proton $Z$ numbers, 
   one can apply the saddle point for large $N$ and $Z$ in a 
  nuclear system.  The effects of the pairing correlations in
   the Fermi system will be worked out
  separately,  see examples in Ref.~\refcite{PRC}.
  Notice that 
  other semi-analytical methods were suggested in the literature
  \cite{JB75,BJ76,PG07}
  to overcome divergence
  of the full saddle point method 
  for low
excitation-energy limit $U \rightarrow 0$.  

 More  general microscopic formulation of the energy level density 
 for mesoscopic systems, in particular for nuclei,  which removes
 the singularity at small excitation energies, is discussed 
in Ref.~\refcite{ZH19}, see also references therein.  
One of the microscopic ways for accounting
     for inter-particle interactions beyond the mean field (shell model)
     in the level density calculations
     was suggested within the Monte-Carlo Shell Model\cite{Or97,AB03,AB15}.
     Another successful 
     approach for taking into
     account the inter-particle
         interactions above the simple shell model
      is given by
     the moments 
     method\cite{KZ16,Ze16,ZK18,KZ20,Ze96}. The main ideas are based on
     the random matrix theory, see Refs.~\refcite{Po65,Ze96,Ze16,Me04}.

For a semiclassical formulation of the 
unified microscopic 
canonical 
and macroscopic grand-canonical 
approximation (shortly, MMA) to the level density, 
we will 
derive a simple non-singular analytical
approximation for the
level density $\rho$ for neutron-proton asymmetric nuclei.
The MMA approach satisfies
the two well-known limits. 
One of them is the 
Fermi gas asymptote, 
$\rho \propto \exp(S)$,  
for a large entropy $S$.
    The opposite limit for small $S$, or excitation energy $U$, 
is the  combinatorics expansion\cite{St58,Er60,Ig72} in powers of $S^2$. 
 For small excitation energies, the empiric formula,
    $\rho\propto \exp[(U-E_0)/T]$, with free parameters $E_0$,  $T$, and
  a  pre-exponent factor
  was suggested for the description of the level density
  of the excited low
  energy states 
  in
Ref.\ \refcite{GC65}.
Later, this formula was named 
as 
constant
``temperature'' model, 
see also Refs.~\refcite{ZK18,ZH19,KZ20}.
The ``temperature'' was considered as an
``effective temperature'' which is
related to the excitation energy
because 
of no direct physical meaning
of temperature for low energy states. 
Following the development of
    Refs.~\refcite{NPA,PRC} we will show below that 
the MMA
has the same power  expansion as the constant
``temperature'' model 
for low energy states
at small excitation energies
$U$.

Such a MMA for  the
level density $\rho$ was 
suggested in Ref.~\refcite{KM79},
 within 
the Strutinsky shell correction method\cite{St67,BD72,BK72} 
based on the Landau-Migdal
quasiparticle theory called as the Finite Fermi System
Theory\cite{La58,AK59,MI67,HS82}.
A mean field potential is used for
calculations of the
energy shell corrections, $\delta E$. The total nuclear energy, $E$,  is the
sum
of these corrections and smooth macroscopic liquid-drop component\cite{MS69}
which can be well approximated by the extended Thomas-Fermi approach
\cite{BG85,BB03}. Thus, 
    within the semiclassical approximation to the
    Strutinsky shell correction method,
the interactions between particles, averaged
over particle numbers, 
    i.e. over many-body microscopic quantum states 
    in realistic
nuclei, 
are approximately taken into
account through the  
extended Thomas-Fermi component beyond the mean field.
Neglecting small {\it residual}-interaction corrections (see \ref{appA}) 
    beyond  
     the macroscopic extended Thomas-Fermi  approach and Strutinsky's
        shell corrections,
one can present\cite{KM79} 
the level density $\rho$  
in terms of the
modified Bessel function
of the entropy variable in the case of
small
 thermal excitation energy $U$ as compared to the
 rotational energy.

         In order to simplify more
         the level density calculations for large particle
    numbers, and for a deeper understanding of the 
correspondence between the classical and the quantum approach,
it is worthwhile to analyze the shell 
effects in the level density $\rho$
 (see Refs.~\refcite{Ig83,So90}) 
and in the entropy, $S$, using 
 the semiclassical
periodic-orbit 
theory 
\cite{SM76,SM77,BB03,MY11}.
This theory, based on the semiclassical
time-dependent propagator, allows
obtaining the total level-density, energy,
free-energy, and grand canonical 
potential in terms of the smooth extended Thomas-Fermi 
term and periodic-orbit
correction.

The MMA
approach \cite{KM79}
was extended \cite{NPA,PRC}
for the description of shell and rotational effects 
on the level density 
itself for
larger 
excitation energies $U$ 
in 
 one-component nucleon systems.
We will 
develop  now this MMA for applications to 
isotopic asymmetric nuclei but for the total level
    density $\rho(E,N,Z)$,
keeping the extension to its spin dependence for a future work.
The level
density parameter $a$ 
 for asymmetric neutron-proton nuclear system
is one of the key quantities
under 
intensive experimental and theoretical 
 investigations\cite{Er60,GC65,BM67,Ig83,Sh92,So90,EB09,ZH19,KS20}. 
As mean values of $a$ are largely
proportional to
the total particle number $A$,
the inverse level density parameter
$K=A/a$ is conveniently introduced
to exclude 
a basic mean $A$-dependence in $a$.
Smooth properties of this 
function of the nucleon number $A$
have been studied
within the framework of the
self-consistent extended Thomas-Fermi 
approach\cite{Sh92,KS18}, see also the
study of shell effects in
 one-component nucleon systems in Refs.~\refcite{NPA,PRC}.
 However,  
the statistical
level density for neutron-proton asymmetric nuclei
is still 
an attractive subject. For instance, 
within the Strutinsky's shell correction approach\cite{BD72},
the major shell
effects in the distribution of single-particle (quasiparticle) 
states near
the Fermi surface 
are quite different for neutron and protons of nuclei,
especially for nuclei far from the $\beta$-stability line.
The present work is concentrated on 
low energy states of
nuclear excitation-energy spectra  below 
the neutron resonances 
for large chains of the nuclear
isotopes.

The structure of the paper is the following.
The level density $\rho(E,N,Z)$ is derived
within the MMA 
by using the periodic-orbit theory 
in Sec. \ref{sec-levden}.
The general shell and isotopic asymmetry
effects (Subsections~\ref{subsec-GP} and
\ref{subsec-jac}) are first discussed within the standard saddle-point method
 asymptote (Subsection~\ref{subsec-SPM}).
We then extend 
the standard 
saddle-point method to a more general MMA approach for describing
the analytical transition from large to small  
excitation energies $U$,
taking 
essentially into account the shell and isotopic
asymmetry effects (Subsection~\ref{subsec-MMA}).
In Section \ref{sec-disc}, we compare our analytical  MMA results for the level
density $\rho(E,N,Z)$, and the inverse level density parameter $K$
with experimental
data for a large isotope chain 
as typical examples of heavy isotopic 
asymmetric nuclei.
Some details of the Landau-Migdal theory and Strutinsky shell correction method, the
periodic-orbit theory, and of the sample method for
extraction of the experimental data from nuclear excitation spectra
are presented in  Appendices A, B and C, respectively.

 \section{Isotopic asymmetric microscopic-macroscopic approach }
 \l{sec-levden}

 \subsection{General points}
 \l{subsec-GP}
 For 
a statistical
description of the level density of a nucleus in
  terms of the conservation  variables; 
the total energy, $E$; and neutron, $N$, and proton, $Z$, numbers; 
one
can begin with
the micro-canonical expression for the level density,
\be\l{dendef1}
\rho(E,N,Z)=
\sum\limits_i\!\delta(E-E_i)~\delta(N-N_i)~\delta(Z-Z_i)
\equiv
\int \frac{\d \beta \d \alpha_n \d \alpha_p}{(2\pi i)^3}~e^{S}~.
\ee
Here, $E_i$, $N_i$, and $Z_i$ represent the system spectrum, and
\be\l{entnp}
S(\beta,\alpha_n,\alpha_p)=\ln \mathcal{Z}(\beta,\alpha_n,\alpha_p)
+\beta E -\alpha_n N -\alpha_p Z
=\beta(E-\Omega-\lambda_nN-\lambda_pZ)~,
\ee
where $\alpha_\tau=\lambda_\tau\beta$,  with $\tau=\{n,p\}$ 
being the isotope subscript.
The entropy $S$, partition $\mathcal{Z}$, and
potential $\Omega$ functions 
are considered 
for arbitrary values of arguments $\beta$ and  $\alpha_\tau$, and
$\Omega=-\ln \mathcal{Z}/\beta$~. 
The integral on right hand side 
of Eq.~(\ref{dendef1}) is the
standard inverse Laplace transformation of the partition function $\mathcal{Z}$.
For large excitation energies,
when the saddle points of the integrals (Eq.~(\ref{dendef1})) over all
variables $\beta$, $\alpha_n$, and $\alpha_p$ exist \cite{Ig83,So90},
we have the standard 
    entropy $S$, partition function $\mathcal{Z}$ and thermodynamic
potential $\Omega$. 
 In the mean field of the Strutinsky's shell correction method\cite{BD72},  
 the single-particle (quasiparticle) 
 level density, $g(\varepsilon)$,
can be also written \cite{BM67} as a sum of the neutron and proton components,
$g=g_n+g_p$~. This leads to a similar isotopic decomposition for the 
potential $\Omega$, 
$\Omega 
=\Omega_n+\Omega_p$, 
where $\Omega_\tau$ is given by 
\be\l{OmFnp}
\Omega_\tau\approx -\beta^{-1} 
\int\limits_0^\infty\d\varepsilon~
g_\tau(\varepsilon)~
\ln\left\{1+\exp\left[\beta\left(\lambda_\tau-
  \varepsilon\right)\right]\right\}~.
\ee
The single-particle 
level density, $g_{\tau}(\varepsilon)$, within
the Strutinsky's shell-correction method\cite{BD72}, 
is a sum of
the statistically averaged smooth, $\tilde{g}_\tau(\varepsilon)$, component,
and the oscillating shell,
$\delta g_{\tau}(\varepsilon)$, 
correction, slightly averaged over the 
single-particle energies,
\be\l{gdecomp}
g_\tau(\varepsilon)\cong \tilde{g}_\tau(\varepsilon)+
\delta g_\tau(\varepsilon)~. 
\ee
Within the semiclassical 
periodic-orbit theory\cite{SM76,SM77,BB03,PRC} (\ref{appB}),
the smooth and
oscillating parts
of the 
level density, $g_\tau(\varepsilon)$, Eq.~(\ref{gdecomp}),
can be approximated, with good accuracy, by
the extended Thomas-Fermi 
level density,
$\tilde{g}_\tau \approx g^{(\tau)}_{\rm \tt{ETF}}$,
and the 
periodic-orbit contribution, $\delta g_\tau\approx
\delta g^{(\tau)}_{\rm scl} $, respectively, see 
Eq.~(\ref{goscsemnp}). 
Using 
  the  
periodic-orbit theory decomposition, Eq.~(\ref{gdecomp}), and Eq.~(\ref{OmFnp}),
one finds 
 for $\tilde{\Omega}_\tau\approx \Omega^{(\tau)}_{\rm\tt{ETF}}$
 the result\cite{KM79,KS20}
\be\l{TFpotF}
\tilde{\Omega}_\tau
=\tilde{E}_\tau 
-\lambda_\tau A_\tau
-\frac{\pi^2}{6\beta^2}\tilde{g}_\tau(\lambda_\tau)~,\quad A_\tau=\{N,Z\}~.
\ee
 Here, $\tilde{E}_\tau\approx E^{(\tau)}_{\rm\tt{ETF}}$ is the
 nuclear extended Thomas-Fermi 
 energy 
component (or the corresponding liquid-drop
energy), and
$\lambda_\tau \approx \tilde{\lambda}_\tau$ is of the 
smooth chemical potential for neutron ($n$) and proton ($p$) systems in the
shell correction method.
With the help of the periodic-orbit 
theory\cite{SM76,SM77,BB03,PRC}, one
obtains\cite{KM79} for the oscillating (shell)
component, $\delta \Omega_\tau$, Eq.~(\ref{OmFnp}),
\be\l{potoscparFnp}
\delta \Omega_\tau= -\beta^{-1} 
\int\limits_0^\infty\d\varepsilon~
\delta g_\tau(\varepsilon)~
\ln\left\{1+\exp\left[\beta\left(\lambda_\tau-
  \varepsilon\right)\right]\right\}
\cong
\delta \Omega^{(\tau)}_{\rm scl}
=\delta F^{(\tau)}_{\rm scl}.
\ee
For the
semiclassical free-energy shell correction, $\delta F^{(\tau)}_{\rm scl}$,
or $\delta \Omega^{(\tau)}_{\rm scl}$,
we incorporate 
the 
periodic-orbit expression\cite{KM79,BB03}: 
\be\l{FESCFnp}
\delta F^{(\tau)}_{\rm scl} \cong \sum^{}_{\rm PO} F^{(\tau)}_{\rm PO}~,\quad\mbox{with}
\quad
F^{(\tau)}_{\rm PO}= E^{(\tau)}_{\rm PO}~
\frac{x^{(\tau)}_{\rm PO}}{
  \sinh\left(x^{(\tau)}_{\rm PO}\right)}~,\quad x^{(\tau)}_{\rm PO}=
\frac{\pi t^{(\tau)}_{\rm PO}}{\hbar \beta}~,
\ee
where $E^{(\tau)}_{\rm PO}$ is a 
periodic-orbit component of the semiclassical 
shell correction energy\cite{SM76},
\be\l{dEPO0Fnp}
 \delta E^{(\tau)}_{\rm scl}=
\sum^{}_{\rm PO}E^{(\tau)}_{\rm PO}~,\qquad E^{(\tau)}_{\rm PO}=\frac{\hbar^2}{(t^{(\tau)}_{\rm PO})^2}\,
g^{(\tau)}_{\rm PO}(\lambda_\tau)~. 
\ee
Here, $t^{(\tau)}_{\rm PO} = {\tt M}~t^{{\tt M}=1}_{\rm PO}(\lambda_\tau)$
is the period 
of particle motion
along 
a periodic  orbit 
(taking into
account its repetition, or period number ${\tt M}$), 
and
$t^{{\tt M}=1}_{\rm PO}(\lambda_\tau)$ is the
period of the neutron ($n$) or proton ($p$) motion along the
primitive 
(${\tt M}=1$) 
periodic orbit in the corresponding $\tau$ potential well 
    with the same
radius, $R=r_0A^{1/3}$.
The period $t^{(\tau)}_{\rm PO}$ (and $t^{{\tt M}=1}_{\rm PO}$), and
the partial oscillating level density component, $g^{(\tau)}_{\rm PO}$,
are taken at the chemical potential, $\varepsilon=\lambda_\tau$,
see also Eq.~(\ref{goscsemnp}) 
for the semiclassical 
single-particle level-density shell correction
(\ref{appB} and Refs.~\refcite{SM76,BB03}). The semiclassical expressions,
Eqs.~(\ref{TFpotF}) and (\ref{potoscparFnp}), are valid for a large
relative action, $\mathcal{S}^{(\tau)}_{\rm PO}/\hbar \sim A^{1/3} \gg 1$~.

Then, expanding 
$x^{(\tau)}_{\rm PO}/\sinh(x^{(\tau)}_{\rm PO})$,
Eq.~(\ref{FESCFnp}), in the shell correction $\delta \Omega_\tau$
(Eqs.~(\ref{potoscparFnp}) and (\ref{FESCFnp})) 
in powers of  $1/\beta^2$
up to the quadratic terms, $\propto 1/\beta^2$, one obtains
\be\l{OmadFnp}
\Omega_\tau \approx E^{(\tau)}_0-\lambda_\tau A_\tau-\frac{a_\tau}{\beta^2}~, 
\ee
where $E^{(\tau)}_0$  is the neutron, or proton ground 
state energy, 
$E^{(\tau)}_0=\tilde{E}_\tau+\delta E_\tau$,
and $\delta E_\tau$ is the shell correction energy of the
corresponding cold system,
$\delta E_\tau \approx \delta E^{(\tau)}_{\rm scl}$
(see Eq.~(\ref{dEPO0Fnp}) and \ref{appB}).
In Eq.~(\ref{OmadFnp}),
$a_\tau$
is the level density parameter 
 with  
    decomposition similar to Eq.~(\ref{gdecomp}):
\be\l{denparnp}
a_\tau=\frac{\pi^2}{6}~g_\tau=\tilde{a}_\tau+\delta a_\tau~,
\ee
where $\tilde{a}_\tau$ is the extended Thomas-Fermi 
component and
$\delta a_\tau$ is the periodic-orbit
shell correction
\be\l{daFnp}
\tilde{a}_\tau 
\approx \frac{\pi^2}{6} g^{(\tau)}_{\rm \tt{ETF}}(\lambda_\tau), \quad
\delta a_\tau \approx 
\frac{\pi^2}{6}\delta g^{(\tau)}_{\rm scl}(\lambda_\tau)~.
\ee
For the extended Thomas-Fermi 
component\cite{BG85,BB03,KS18,KS20}, $g^{(\tau)}_{\rm \tt{ETF}}$,
one takes 
into account the
self-consistency with Skyrme forces\cite{AS05}. 
For
the semiclassical 
periodic-orbit
level-density shell corrections\cite{SM76,SM77,BB03,MY11,PRC,PRC},
$\delta g^{(\tau)}_{\rm scl}(\lambda_\tau)$, we use
Eq.~(\ref{goscsemnp}).

Expanding the entropy, Eq.~(\ref{entnp}), over the Lagrange multipliers
$\alpha_\tau$ near the saddle points $\alpha^\ast_\tau$, 
 one can use the
saddle point equations (particle number conservation equations),
\begin{equation}\label{Seqsdnp}
\left(\frac{\partial S}{\partial \alpha_\tau}\right)^\ast\equiv
\left(\frac{\partial \ln Z}{\partial \alpha_\tau}\right)^\ast-A_\tau=0~.
\end{equation}
Integrating then over $\alpha_\tau$ in Eq.~(\ref{dendef1}), by the standard saddle-point method, 
one obtains
\be\l{rhoE1Fnp}
\rho(E,N,Z) \approx 
\frac{1}{4\pi^2 i}
\int \d \beta~\beta
\mathcal{J}^{-1/2}
\exp\left(\beta U + a/\beta\right)~,
\ee
where $U=E-E_0$ is the excitation energy ($E_0=E^{(n)}_0+E^{(p)}_0$),
 and   
\be\l{anp}
a=a_n+a_p, \qquad g=g_n+g_p~, 
\ee
and $a_\tau$ is
the $\tau$ component of the 
level density parameter, 
given by Eqs.~(\ref{denparnp}), and (\ref{daFnp}). In 
equation (\ref{rhoE1Fnp}),
$\mathcal{J}$ is the two-dimensional 
 Jacobian determinant, 
$\mathcal{J}(\lambda^\ast_n,\lambda^\ast_p)$, taken 
 at the saddle point, 
 $\lambda_\tau=\lambda^\ast_\tau=\alpha^\ast_\tau/\beta$,
Eq.~(\ref{Seqsdnp}), 
at a given $\beta$,
\be\l{Jacnp}
\mathcal{J}=\mathcal{J}\left(
      \frac{\partial \Omega}{\partial \lambda_n},
      \frac{\partial \Omega}{\partial \lambda_p};
      \lambda_n,\lambda_p\right)^\ast =
      \left(\frac{\partial^2 \Omega}{\partial \lambda_n^2}
      \frac{\partial^2 \Omega}{\partial \lambda_p^2}\right)^\ast~,
\ee
where the asterisk indicates
the saddle point 
for the integration over $\alpha_\tau$ at any $\beta$. 
    In the following, for simplicity of
notations, we will omit the asterisk at $\lambda^\ast$.

\subsection{Jacobian calculations}
\l{subsec-jac}

Taking 
the derivatives of Eq.~(\ref{OmadFnp}) for the potential $\Omega$
 with respect to $\lambda_\tau$,
in the Jacobian $\mathcal{J}$, Eq.~(\ref{Jacnp}),
up to linear terms in expansion over $1/\beta^2$ (Ref.~\refcite{PRC}), one
obtains\footnote{We shall present the Jacobian calculations for the main case
  of $\delta g<0$
  near the minimum of the level density and 
  shell correction energy, as mainly applied below.
  For the case of a
  positive $\delta g$ we change, for convenience, signs so that we will get
  $\xi>0$.}
\be\l{Jacnp2}
\mathcal{J} \approx \mathcal{J}_0 + \mathcal{J}_2/\beta^2=
\mathcal{J}_0 \left(1+\xi\right),
\ee
where
\be\l{xiparnp}
\xi=\overline{\xi}/\beta^2,\qquad
\overline{\xi}=\mathcal{J}_2/\mathcal{J}_0~,
\ee
 with 
\be\l{J12np}
\mathcal{J}_0=
g_n(\lambda_n)g_p(\lambda_p),\quad
\mathcal{J}_2=
a^{\prime\prime}_pg_n(\lambda_n)+a^{\prime\prime}_ng_p(\lambda_p)~,
\ee
see Eqs.~(\ref{Jacnp}), and (\ref{OmadFnp}).
 Up to a small asymmetry parameter squared, $~~X^2=$\\ $(N-Z)^2/A^2$, one has
approximately, 
$\lambda_n=\lambda_p=\lambda$,
and $~g_ng_p=g^2/4~$ (see Eq.~(\ref{anp}) for $g$). 
Then, correspondingly, one can simplify
Eqs.~(\ref{J12np}) with (\ref{denparnp}) 
to have
\be\l{J02snp}
\mathcal{J}_0=
\frac14 g^2,\quad
\mathcal{J}_2=
\frac{\pi^2}{12}g^{\prime\prime}g~.
\ee

According to Eq.~(\ref{gdecomp}), a
decomposition of the Jacobian, Eq.~(\ref{Jacnp2}),
in terms of its 
smooth extended Thomas-Fermi 
and linear oscillating periodic-orbit 
components 
    of $\mathcal{J}_0$ and $\mathcal{J}_2$
can be found straightforwardly
with the help of 
Eqs.(\ref{J12np}) and  (\ref{denparnp})
(see also Ref.~\refcite{PRC}),
\be\l{Jdecomp}
\mathcal{J}_0=\tilde{\mathcal{J}}_0+\delta \mathcal{J}_0, \qquad 
\mathcal{J}_2=\tilde{\mathcal{J}}_2+\delta \mathcal{J}_2\approx \delta \mathcal{J}_2~.
\ee
 As demonstrated in \ref{appB}, the dominance
of derivatives of the semiclassical expression (\ref{goscsemnp}) for
the single-particle 
level density
shell corrections,
$\delta g_{\rm scl}$, in Eq.~(\ref{J02snp}) for
$\delta \mathcal{J}^{}_2$, 
led to the
last approximation in Eq.~(\ref{Jdecomp}).
For smooth,
$\tilde{\mathcal{J}}$, and oscillating,
$\delta \mathcal{J}$,  components of
$\mathcal{J}\cong \tilde{\mathcal{J}}+\delta \mathcal{J}$, Eq.~(\ref{J12np}),
one
finds with
the help of 
Eq.~(\ref{daFnp}),
\be\l{tilJnp}
\tilde{\mathcal{J}}\approx \tilde{\mathcal{J}}_0=
\tilde{g}_{n}\tilde{g}_{p}~,\quad
\delta \mathcal{J}= \delta \mathcal{J}_0 +\delta \mathcal{J}_2/\beta^2,
\ee
where $\tilde{g}_{\tau}\approx
g^{(\tau)}_{\rm{\tt{ETF}}}(\lambda_\tau)$ is approximately the 
(extended) Thomas-Fermi $\tau$-level-density component.
For linearized oscillating  major-shells components of $\delta \mathcal{J}$, 
one approximately arrives at
\be\l{dJnp}
\delta \mathcal{J}_0\approx 
\tilde{g}_{n}\delta g_p
    +\tilde{g}_{p}\delta g_n~,\quad
\delta \mathcal{J}_2\approx 
 - \frac{2\pi^4}{3}\left[
    \frac{\tilde{g}_{n} \delta g_p}{
      \mathcal{D}_{p}^2}
    +\frac{\tilde{g}_{p} \delta g_n}{
      \mathcal{D}_{n}^{2}}
    \right]~,
  \ee
  where $\delta g_\tau$ is the 
  periodic-orbit shell component,
  $\delta g_\tau\approx \delta g^{(\tau)}_{\rm scl}$
  (see Eq.~(\ref{goscsemnp})),
   $\mathcal{D}_{\tau} = \mathcal{D}^{(\tau)}_{\rm sh}=\lambda_\tau/A^{1/3}$ is
  approximately the distance between major (neutron or proton) shells
  given by
  Eq.~(\ref{periodenp}). 
   Again, up to  terms of the order of $X^2$,
   one simply finds from Eqs.~(\ref{tilJnp}) and (\ref{dJnp}),
    \be\l{J02dcompsnp}
    \tilde{\mathcal{J}}\approx \frac14\tilde{g}^2,\quad
   \delta \mathcal{J}_0\approx 
   \frac12\tilde{g}\delta g~,\quad
   \delta \mathcal{J}_2\approx
     -\frac{\pi^4}{3}\frac{g\delta g}{\mathcal{D}^2}~,
     \ee
where $\mathcal{D}=\lambda/A^{1/3}$.
Note that for
thermal excitations smaller or of the order of those of neutron resonances,
 the main contributions of the oscillating potential, 
$\delta \Omega_\tau$, and Jacobian, $\delta \mathcal{J}$, components
as functions of $\lambda_\tau$,
 are coming from the
differentiation of the  sine in
the periodic-orbit 
level density  
component, $g^{(\tau)}_{\rm PO}(\lambda_\tau)$, Eq.~(\ref{goscsemnp}),
 through
 the 
 periodic-orbit action phase
$\mathcal{S}^{(\tau)}_{\rm PO}(\lambda_\tau)/\hbar$.
The reason is that, 
for large 
particle numbers, $A$, the
semiclassically large parameter,
$\sim\mathcal{S}^{(\tau)}_{\rm PO}/\hbar \sim A^{1/3}$, 
    leads to 
dominating contribution, much larger than that coming
from  differentiation of other terms, such as,
the $\beta$-dependent function
$x^{(\tau)}_{\rm PO}(\beta)$, and 
the periodic-orbit 
period $t^{(\tau)}_{\rm PO}(\lambda)$.
Thus,  in the linear approximation over $1/\beta^2$,
we simply arrive 
to Eq.~(\ref{dJnp}), similarly to the derivations in 
    Ref.~\refcite{PRC}.

    In the linear approximation in $1/\beta^2$, one finds
    from Eq.~(\ref{xiparnp}) for $\xi$
and
Eq.~(\ref{FESCFnp}) for $x^{(\tau)}_{\rm PO}$, see also Eqs.~(\ref{tilJnp}), 
(\ref{dJnp}) and (\ref{anp}),
\be\l{xib}
\overline{\xi}
\approx \delta \mathcal{J}_2/\tilde{\mathcal{J}_0}\approx
\overline{\xi}_n
+\overline{\xi}_p~,\qquad \overline{\xi}_\tau=-\frac{2\pi^4 \delta g_{\tau}}{3\tilde{g}_\tau\mathcal{D}^2_\tau}~,
  \ee 
 see also 
 Eq.~(\ref{periodenp}) for $D^{}_\tau$.
 For convenience, 
 introducing 
 the dimensionless 
 shell
  correction energy,
$\mathcal{E}^{(\tau)}_{\rm sh}$, 
  in units of   
  the smooth extended Thomas-Fermi 
  energy per particle, $E^{(\tau)}_{\rm \tt{ETF}}/A$,
      one can present
      Eq.~(\ref{xib}) (for $\delta E<0$) as:
\be\l{xibdEnp}
\overline{\xi}_\tau
\approx \frac{4 \pi^6A^{1/3} \mathcal{E}^{(\tau)}_{\rm sh}}{3\lambda^2_\tau}~,
\quad 
\mathcal{E}^{(\tau)}_{\rm sh}=-\frac{A\delta E_\tau}{E^{(\tau)}_{\rm \tt{ETF}}}~.
\ee
The smooth extended Thomas-Fermi 
energy $E^{(\tau)}_{\rm \tt{ETF}}$
       can be 
    approximated as 
    $E^{(\tau)}_{\rm \tt{ETF}}\approx
    \tilde{g}_\tau(\lambda_\tau)\lambda_\tau^2/2 $~. 
    The 
    shell correction energy, $\delta E=\delta E_n+\delta E_p$,
 is  expressed,  for a major shell structure with
 semiclassical accuracy, through the periodic-orbit 
sum\cite{SM76,SM77,BB03,MY11,PRC}
      in Eq.~(\ref{dEPO0Fnp}),
  $\delta E_\tau\approx \delta E^{(\tau)}_{\rm scl}$, where 
  (see \ref{appB}) 
  \be\l{dedgnp}
  \delta E^{(\tau)}_{\rm scl} 
  \approx
  \left(\frac{D_{\tau}}{2 \pi}\right)^2~ \delta g_{\tau}(\lambda_\tau)~.
  \ee

    The correction, $\propto 1/\beta^4$, of
    the expansion in  $\propto 1/\beta^2$ of both 
        the potential shell correction, Eq.~(\ref{potoscparFnp}) with
        Eq.~(\ref{FESCFnp}),
    and the Jacobian,
    Eq.~(\ref{Jacnp}),
         through the
    oscillating part, $\delta \mathcal{J}$, 
 is 
 relatively small
 for $\beta $ which, evaluated at the critical saddle point 
 values $T=1/\beta^\ast$, is related to the
  chemical potential $\lambda_\tau$
  as $T \ll \lambda_\tau$. Thus, the temperatures $T=1/\beta^\ast$,
when the saddle point $\beta=\beta^\ast$ exists, 
 are assumed to be 
much smaller than the chemical
potentials $\lambda_\tau$. The high order,
  $\propto 1/\beta^4$, term of this expansion  can be
  neglected under the following condition (subscripts $\tau$
  are omitted for a small asymmetry parameter $(N-Z)^2/A^2$,
  see also
  Ref.~\refcite{PRC}): 
 \be\l{condUnp}
\frac{1}{\tilde{g}}\siml U\ll
\sqrt{\frac{90}{7}}\frac{a\lambda^2}{2\pi^4 A^{2/3}}~.
\ee
Using typical values for
parameters 
$\lambda\approx 40$ MeV, $A=200$, and
$a =A/K\sim 20 $~MeV$^{-1}$ ($K\sim 10$~MeV), $1/\tilde{g}\sim 0.1-0.2$~MeV,
 one finds, 
numerically, that
the r.h.s. of this inequality is of the order of the chemical potential, $\lambda$, 
see Ref.\ \refcite{KS18}. Therefore, one obtains approximately
$U \ll \lambda$.
For simplicity,
          the small shell and temperature corrections to
        $\lambda_\tau(A_\tau)$ obtained  from the conservation
          equations,  Eq.~(\ref{Seqsdnp}),
          can be neglected.
           Using the linear shell correction approximation 
           of the leading order \cite{BD72}
           and constant particle
           number density of symmetric nuclear matter, 
           $\rho^{}_0=2k_F^{3}/3\pi^2=0.16$ fm$^{-3}$
           ($k^{{n}}_F\approx k^{{p}}_F\approx k_F=1.37$ fm$^{-1}$
           is the Fermi momentum in units of $\hbar$),
           one finds about a constant value for 
           the chemical potential,
           $\lambda_\tau \approx \hbar^2k^{2}_F/2m \approx 40$ MeV,
           where $m$ is the nucleon mass.
       In the derivations of the condition (\ref{condUnp}),
       we used the periodic-orbit theory  
       distances between major shells,
       $D^{(\tau)}_{\rm sh}$, Eq.~(\ref{periodenp}),
       $D^{(\tau)}_{\rm sh}\approx \lambda/A^{1/3}$.
Evaluation of the upper limit for the
excitation energy at the saddle point $\beta=\beta^\ast=1/T$
is justified because: this upper
limit is always so large that this point does  certainly exist.
Therefore, for 
 consistency, one can neglect the
  quadratic, $1/\beta^2$ (temperature $T^2$), corrections to the Fermi
  energies $\varepsilon^{(\tau)}_{F}$ in the chemical
  potential\footnote{In
    our semiclassical picture,
           it is convenient to determine  
          the Fermi energy,  $\varepsilon^{(\tau)}_{F}$,
          from 
          the depths of the neutrons and protons potentials, which
          are different due to Coulomb interaction.
    Note that the upper energy  levels in the neutron and proton
    potential wells are approximately the same near
    the $\beta$ stability line.},
  $\lambda_\tau\approx \varepsilon^{(\tau)}_{F}$ (or,  
     $\lambda_\tau \approx \lambda \approx \varepsilon^{}_{F} $), for large 
  particle numbers $A$ and small asymmetry parameter $X^2$.

  \subsection{Shell and isotopic asymmetry effects within the
    saddle-point method} 
\l{subsec-SPM}

For simplicity, one can start with a  
direct application of the standard 
saddle-point approach
for calculations of the inverse Laplace integral over $\beta$ in Eq.~(\ref{rhoE1Fnp}).
In this 
way, including
    the shell (Ref.~\refcite{PRC}) and isotopic asymmetry effects,
      one arrives at 
\be\l{SPMgennp}
\rho(E,N,Z)\approx
\frac{a^{3/4}\exp(2\sqrt{aU})}{4\pi~U^{5/4}~\sqrt{\pi\mathcal{J}_0(1+\xi^\ast)}}
\approx
\frac{\sqrt{\pi}\exp(2\sqrt{aU})}{12 a^{1/4}~U^{5/4}~\sqrt{1+\xi^\ast}}~.
\ee
 Here, $a=\pi^2 g/6$ is the total level density parameter,
    Eqs.~(\ref{anp}) 
and (\ref{daFnp}), $\mathcal{J}_0$ is the component of the
Jacobian $\mathcal{J}$,
Eq.~(\ref{Jacnp2}), which is independent of
 $\beta$ but depends on the shell structure, 
  and
\be\l{pars}
\xi^\ast=\frac{\overline{\xi}}{\beta^{\ast\;2}}
\approx\frac{2\pi^4 U }{3a A^{1/3}}\mathcal{E}_{\rm sh}^{\prime\prime}(\lambda)
  \approx\frac{8\pi^6 U A^{1/3}}{3a\lambda^2}\mathcal{E}_{\rm sh}~.
\ee
 The relative shell correction,
    $\mathcal{E}_{\rm sh} \approx \mathcal{E}_{\rm sh}^{(\tau)}$, given
    by Eq.~(\ref{xibdEnp}), 
    is almost independent of $\tau$ for small $X^2$,
see also Eqs.~(\ref{xib}) and (\ref{dedgnp}). 
 The asterisk means $\beta=\beta^\ast=\sqrt{a/U}$ at the saddle point.
 In the second equation of (\ref{SPMgennp}), and of (\ref{pars})
 we used
    $\lambda_n\approx \lambda_p\approx \lambda$
 for a small asymmetry parameter, $X^2$, 
 together with Eq.~(\ref{d2Edl2}) for the
derivatives of the energy shell corrections 
    and Eq.~(\ref{periodenp}) for
the mean distance between
neighboring major shells near the Fermi surface,
$\mathcal{D}\approx \lambda/A^{1/3}$.

In Eq.~(\ref{SPMgennp}), the quantity $\xi^\ast$ 
is $\xi$ in
    Eq.~(\ref{xiparnp}), 
taken at the saddle point,
    $\beta=\beta^\ast$ ($\lambda_\tau=\lambda_\tau^\ast$). 
 This quantity
is the sum of the two $\tau$ contributions, $\xi^\ast=\xi^\ast_n
+\xi^\ast_p$, $~~\xi^\ast_n\approx \xi^\ast_p$.  The value of $\xi^\ast$ 
is
 approximately proportional to the excitation energy, $U$, and
to the  relative 
energy shell corrections, $\mathcal{E}_{\rm sh}$, 
Eq.~(\ref{xibdEnp}), and inversely proportional to the level density
parameter, $~a~$,
 with 
$~\xi^\ast\propto U A^{1/3}\mathcal{E}_{\rm sh}/(a \lambda^2)$. 
 For typical parameters
  $\lambda=40$ MeV, $A\sim 200$,
  and 
  $\mathcal{E}_{\rm sh}=|\delta E~A/E_{\rm \tt{ETF}}|\approx 2.0$
  \cite{BD72,MSIS12},
  one finds the estimates
  $\xi^\ast\sim 0.1 - 10$ for temperatures $T \sim 0.1-1$ MeV. 
      This corresponds approximately
  to a rather wide 
  excitation energies $U=0.2-20$~ MeV for $K=A/a=10$~MeV, see
  Ref.~\refcite{KS18}
  ($U=0.1-10$~MeV for $K=20$~MeV). This energy range includes the 
     low-energy states and states
   significantly
  above the neutron resonances.
  Within the 
  periodic-orbit theory\cite{SM76,BB03,MY11} 
  and 
  extended Thomas-Fermi approach \cite{BG85,BB03,KS20},
    these values are given finally by using the realistic
    smooth energy $E_{\rm \tt{ETF}}$ for which
    the binding energy\cite{MSIS12}   
  is $BE\approx E_{\rm \tt{ETF}}+ \delta E$.

    Eq.~(\ref{SPMgennp})
     is a more general shell-structure Fermi-gas (SFG) 
    asymptote,
    at 
     large excitation energy,
  with respect to the well-known\cite{Be36,Er60,GC65}
   Fermi gas (FG) 
  approximation for $\rho(E,N,Z)$, which is equal to Eq.~(\ref{SPMgennp}) at
  $\xi^\ast \rightarrow 0$,
  \be\l{FG}
\rho(E,N,Z)\rightarrow 
\frac{\sqrt{\pi}\exp(2\sqrt{aU})}{12 a^{1/4}~U^{5/4}}~.
\ee
Notice that a shift of the inverse level density parameter
$K$ due to shell effects with increasing excitation energies
which is related to temperatures of the order of 1-3 MeV is
discussed in Refs.~\refcite{PRC,SN90,SN91}.

\subsection{Shell and isotopic asymmetry effects within the MMA}
\l{subsec-MMA}

Under 
  the condition of Eq.~(\ref{condUnp}),
one can obtain simple analytical expressions for the level density
$\rho(E,N,Z)$, beyond the standard saddle-point method, 
from the integral representation
(\ref{rhoE1Fnp}). 
 The square root
Jacobian factor $\mathcal{J}^{-1/2}$ in its integrand can be 
simplified
very much
by expanding\footnote{At each finite order of these expansions, one can  
accurately take\cite{PRC}
the inverse Laplace transformation.
Convergence of the corresponding corrections to the level density,
Eq.~(\ref{rhoE1Fnp}),
after applying this inverse
transformation,
can
be similarly proved as carried out in Ref.~\refcite{PRC}.} it in small values of $\xi$ or of $1/\xi$
(see Eq.~(\ref{Jacnp2})).
Expanding now this Jacobian factor
    at linear order in $\xi$ and $1/\xi$, 
one arrives at two different approximations marked below by cases
(i) and (ii), respectively. 
 Then, taking 
the
inverse Laplace transformation over 
$\beta$   in Eq.~(\ref{rhoE1Fnp}),
with the transformation of $\beta$ to the inverse variable, $1/\beta$,
more accurately
(beyond the standard saddle-point\cite{PRC}), 
one approximately obtains (see Ref.~\refcite{PRC}),
    \begin{equation}\label{denbesnp}
    \rho \approx \rho^{}_{\tt{MMA}}(S)
    =\overline{\rho}_\nu~f_\nu(S)~,~~~f_\nu(S)=
  S^{-\nu}I_{\nu}(S)~,
  \end{equation}
    with
    the entropy $S$ 
    given by $S=2\sqrt{a U}$,
     where $a$ is the sum of the extended Thomas-Fermi term and
    periodic-orbit shell correction, see Eqs.~(\ref{anp}) and (\ref{denparnp}). 
    For small,
        $\xi \sim \xi^\ast \ll 1$, case (i), and
        large, $\xi \sim \xi^\ast \gg 1$,
        case (ii), where $\xi^\ast$ is, thus,
    the critical shell-structure quantity,
    given by 
    expressions (\ref{pars}),
    $\xi^\ast \propto \mathcal{E}_{\rm sh}$,
             one finds
    $\nu=2$  for case (i) and $3$ for case (ii), 
    respectively.
     In case (i) and (ii),
    called below the MMA1 and MMA2 approaches,  respectively,
    one obtains
    Eq.~(\ref{denbesnp})
    with different coefficients $\overline{\rho}_\nu$ 
    (see also Ref.~\refcite{PRC}),
\bea\l{rho2}
&\rho^{}_{\tt{MMA1}}(S)=\overline{\rho}_{2}S^{-2}I_{2}(S),\qquad
\overline{\rho}_{2}=\frac{2 a^2}{\pi\sqrt{\mathcal{J}_0}}\approx \frac{2\pi a}{3}
\qquad \mbox{(i)},\\
&\rho^{}_{\tt{MMA2}}(S)=\overline{\rho}_{3}S^{-3}I_{3}(S),\qquad
\overline{\rho}_{3}\approx 
\frac{4a^3}{\pi \sqrt{\overline{\xi}\mathcal{J}_0}}
\approx \frac{4\pi a^2}{3\sqrt{\overline{\xi}}} \qquad \mbox{(ii)},
\l{rho3}~
\eea
where $\overline{\xi}$ and $\mathcal{J}_0$ are given by
Eqs.~(\ref{xiparnp}) and
(\ref{J12np}), respectively, see also Eq.~(\ref{J02snp}) for the
last approximations.
For the 
Thomas-Fermi approximation to the coefficient $\overline{\rho}_{3}$ within the
 case (ii) 
one finds\cite{NPA,PRC}
\be\l{rho3-2b}
\rho^{}_{\tt{MMA2b}}(S)=\overline{\rho}^{(2b)}_{3}S^{-3}I_{3}(S),\qquad
\overline{\rho}^{(2b)}_{3}\approx \frac{2 \sqrt{6}~\lambda a^2}{3}~.
\ee
 In the derivation of the coefficient, $\overline{\rho}^{(2b)}_{3}$, we assume in
    Eq.~(\ref{rho3}) for $\overline{\rho}_3$
that the magnitude
of the relative shell corrections
    $\mathcal{E}_{\rm sh}$, $\overline{\xi}\propto \mathcal{E}_{\rm sh}$,
see Eqs.~(\ref{xiparnp}), 
(\ref{J02snp}), and (\ref{xibdEnp}),
are  extremely
    small but their derivatives yield large
    contributions through the 
    level density derivatives  $g^{\prime\prime}(\lambda)$,
$g \propto A/\lambda$, 
  as in the 
 Thomas-Fermi approach.
For large entropy
$S$, one finds from Eq.~(\ref{denbesnp})
 \begin{equation} 
 f_\nu(S) =\frac{\exp(S)}{S^{\nu}\sqrt{2\pi S}}\left[1+\frac{1-4\nu^2}{8S}
    +\mbox{O}\left(\frac{1}{S^2}\right)\right].
 \label{rhoasgennp}
\end{equation}
 The same leading results in the expansion (\ref{rhoasgennp}) 
for $\nu=2$ (i) and  $\nu=3$ (ii) 
at large excitation energies $U$
are also derived 
from the shell-structure Fermi gas  
formula (\ref{SPMgennp}).  
     At small entropy, $S \ll 1$, one obtains
     also from Eq.~(\ref{denbesnp})
     the finite combinatorics power
expansion\cite{St58,Er60,Ig72}: 
\begin{equation}
    f_\nu(S)=
  \frac{2^{-\nu}}{\Gamma(\nu+1)}\left[1+\frac{S^2}{4(\nu+1)}+
    \mbox{O}\left(S^4\right)\right],
\label{den0gennp}
\end{equation}
where $\Gamma(x)$ is the Gamma function.
    This expansion over powers of 
    $S^2 \propto U$ 
    is the same 
    as that of  the 
    constant ``temperature'' model\cite{GC65,ZK18,ZH19,KZ20},    
    used often for the level density calculations,
    but here, as in Ref.~\refcite{PRC}, we have it without
free fitting parameters.
\begin{figure}
  \centerline{\includegraphics*[width=8.0cm]{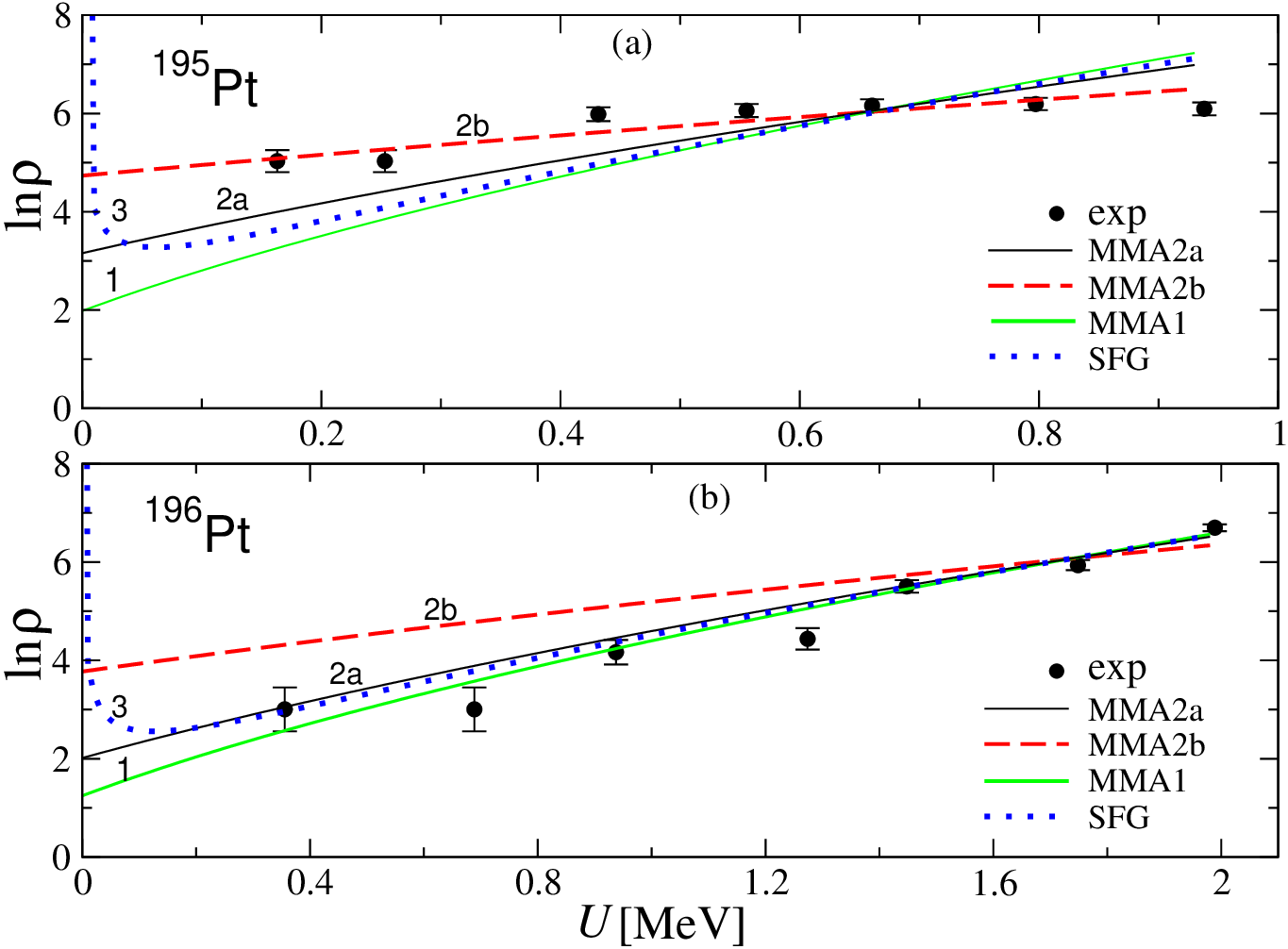}}

 \caption{ 
   Level density, $\mbox{ln}\rho(E,N,Z)$,
    for
   low energy states in nuclei $^{195}$Pt (a),
   and $^{196}$Pt (b)
   were calculated  within different
    approximations: The MMA 
    solid green ``1'' and black ``2a'',
    Eqs.~
    (\ref{rho2})
    and (\ref{rho3})
    at the  realistic relative shell correction\cite{MSIS12}
    $\mathcal{E}_{\rm sh}$, respectively; the MMA dashed red ``2b'',
    Eq.~(\ref{rho3-2b}),
        and
     the generalized shell-structure Fermi-Gas  
 rare blue dots, 
      Eq.~(\ref{SPMgennp}),
      approaches, respectively, are presented.
      The realistic values of $
     \mathcal{E}_{\rm sh}$= 0.77 (a), and 0.84 (b) for MMA2 are    
     taken from Ref.~\protect\refcite{MSIS12}
      (with chemical potentials $\lambda_n\approx\lambda_p\approx
     \lambda$, where
      $\lambda=40$ MeV).
     Experimental dots with error bars
     are obtained 
     from ENSDF database\protect\cite{ENSDFdatabase} 
    by using the sample method (\ref{appC}).
}
\protect\label{fig1}
\end{figure}

 In
contrast to the finite MMA limit (\ref{den0gennp}) for the level density,
Eq.~(\ref{denbesnp}), the asymptotic SFG (Eq.~(\ref{SPMgennp}))
and FG (Eq.~(\ref{FG})) expressions are obviously divergent
at $U\rightarrow 0$.
 Notice also that the MMA1 approximation for the level density,
  $\rho(E,N,Z)$,
  Eq.~(\ref{rho2}),
  can be applied also for
  large excitation energies, $U$, with respect to the
  collective rotational excitations,
  as the SFG and FG approximations 
  if one can neglect shell effects, $\xi^{\ast}\ll 1$.
  Thus, the level density $\rho(E,N,Z)$ in the case (i), Eq.~(\ref{rho2}),
  has wider range of the applicability over the excitation energy variable
  $U $ than the MMA2 case (ii).
  The MMA2 approach has, however, another advantage 
  of describing the important shell structure effects. 
      The main effects of
  the inter-particle
  interaction, statistically averaged over particle numbers,
  beyond the shell correction of the mean field within the Strutinsky's
  shell correction 
      method was taken into account by the extended Thomas-Fermi
  components of MMA expression
  (\ref{denbesnp}) for the level density, $\rho(E,N,Z)$. These components
  are given by 
  the extended Thomas-Fermi potential,
  $\Omega^{}_{\rm \tt{ETF}}$, Eq.~(\ref{TFpotF}),
  and the level-density parameter, $a^{}_{\rm \tt{ETF}}$,
  Eq.~(\ref{daFnp}), counterparts of the corresponding
  total quantities,
  Eqs.~(\ref{OmFnp}) and (\ref{denparnp}).

\section{Discussion of the results}
\l{sec-disc}

Fig.~\ref{fig1} and Table \ref{table-1} 
show
different theoretical 
    (MMA, Eq.~(\ref{denbesnp});  SFG, Eq.~(\ref{SPMgennp}), 
and;  standard FG, Eq.~(\ref{FG}))
approaches
for the statistical level density $\rho(E,N,Z)$ (in logarithms)
as functions of the
excitation energy $U$. 
They are compared to the
experimental
data obtained by the sample method as explained in \ref{appC}.

\begin{figure}
  \centerline{\includegraphics*[width=10.0cm]{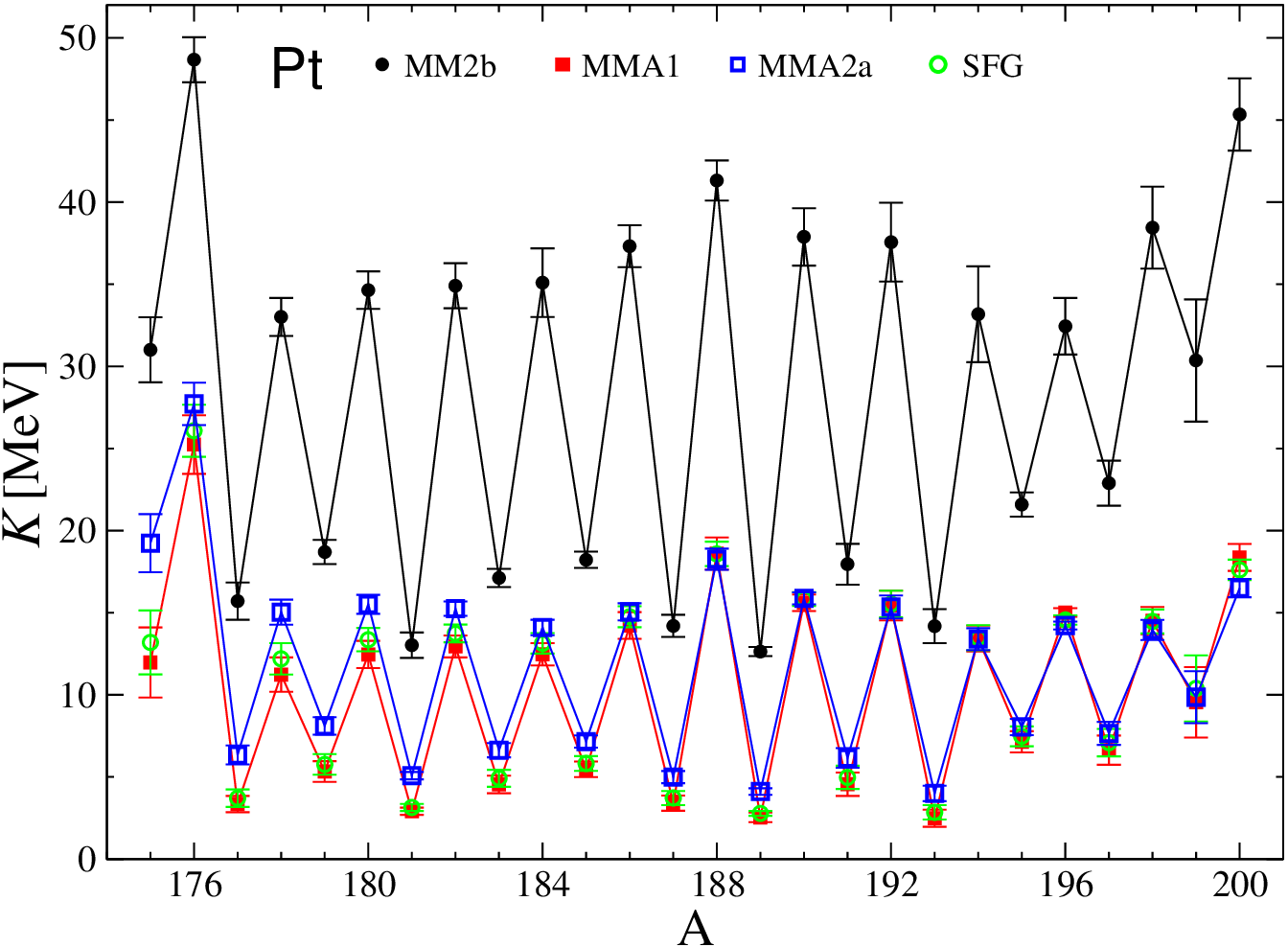}}
     
 \caption{
      Inverse level density parameters $K$ (with errors) 
      for Pt isotopes 
      are shown as
     function of the particle numbers $A$ within a long chain A=175-200.
     Different symbols correspond to several approximations:
      the close black
     dots for MMA2b,
     full red squares for MMA1, larger blue open squares for MMA2a,  and
     green open circles 
     for  SFG.      
}
\label{fig2}
\end{figure}

\begin{figure}
  \centerline{\includegraphics*[width=10.0cm]{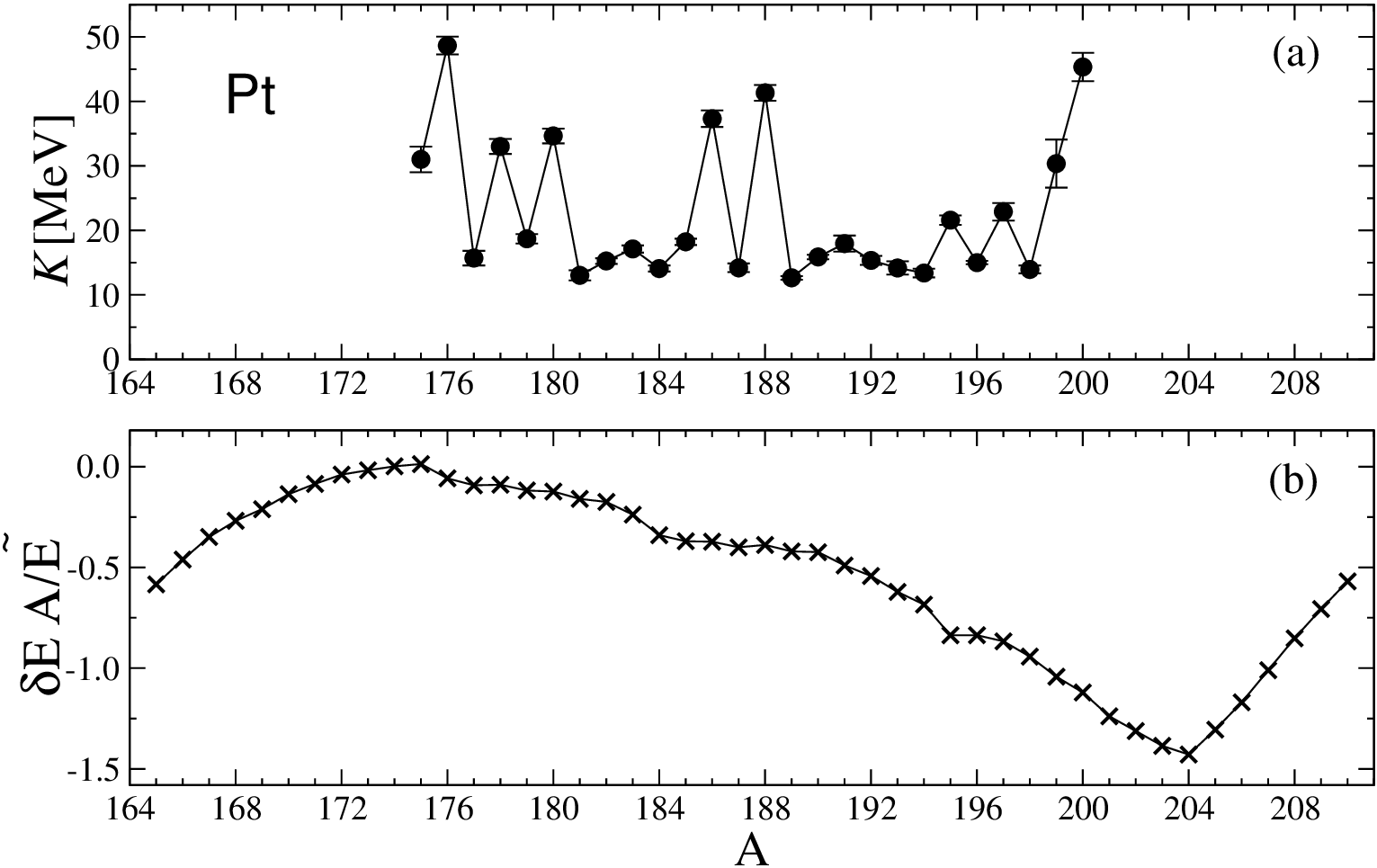}}
      
 \caption{
      Inverse level density parameters $K$ (with errors) 
      for Pt isotopes 
      are shown as
     function of the particle numbers $A$ within a long chain A=175-200.
     (a):  the close black
     dots 
     for the MMA approach
      taken with the smallest relative 
     error parameter
    $\sigma$, Eq.~(\ref{chi}), among all MMAs. (b):  The
         relative shell correction energies,
         $\delta E A/\tilde{E}$ are taken
         from Ref.~\protect\refcite{MSIS12}.
}
\label{fig3}
\end{figure}

The 
level densities 
shown in Fig.~\ref{fig1}
are calculated  
using the inverse level density parameter $K$, 
    see Figs.~\ref{fig2} and \ref{fig3},
found from their least mean-square 
fits 
to the experimental data for
two 
isotopes of platinum, $^{195}$Pt and $^{196}$Pt, as a typical example.
The experimental data shown by dots 
are obtained for the statistical level density $\rho(E,N,Z)$
from
the spectra of nuclear excited states 
by using the sample method\cite{So90} for their distributions 
(see \ref{appC}).

As in Ref.~\refcite{PRC} for isotopes $^{166}$Ho and $^{144}$Sm, Fig.~\ref{fig1}
and Table~\ref{table-1} 
 present 
the two opposite situations
concerning the states distributions as functions of the excitation energy $U$.
We 
show results for the nucleus $^{195}$Pt (a)
with a large number of the low energy states 
below excitation energy of about
1 MeV. 
For $^{196}$Pt (b)
one has 
very small number of low energy states 
below 
    the same energy of about 1 MeV 
 (see ENSDF database\cite{ENSDFdatabase} 
    and Table \ref{table-1} for maximal excitation energies
 $U_{\rm max}$).
But there are many states in $^{196}$Pt with 
 excited energy of
above 1 MeV up to essentially 
larger
    excitation energy of about 2 MeV. 
    According to Ref.~\refcite{MSIS12},
the shell effects, measured by
$\mathcal{E}_{\rm sh}$, Eq.~(\ref{xibdEnp}), are significant in both
these deformed nuclei,
see Fig.~\ref{fig3}(b).

In 
Fig.~\ref{fig1}, 
the results of the 
MMA1 and MMA2 approaches, 
Eqs.~(\ref{rho2}) and (\ref{rho3}), respectively,
are compared with the  SFG approach, Eq.~(\ref{SPMgennp}),
with a focus on shell
effects. 
The SFG results are
very close
    to those of the well-known FG asymptote, Eq.~(\ref{FG}),
     which neglects
the shell effects,
see also Table \ref{table-1}.
The results of the MMA2a approach, Eq.~(\ref{rho3}), 
in the dominating 
    shell effects case (ii)
    ($\xi^\ast \gg 1$, Eq.~(\ref{pars}))
with 
the realistic relative shell correction,
$~\mathcal{E}_{\rm sh}~$ (Ref.~\refcite{MSIS12}),
 are shown 
versus those of  a  small 
shell effects approach MMA1 (i), 
 Eq.~(\ref{rho2}), valid at
        $~\xi^\ast \ll 1~$. 
The results of the
limit of the MMA2 to a very small 
value of $~\mathcal{E}_{\rm sh}$,~ but still within the case (ii),
Eq.~(\ref{rho3-2b}),
called as MMA2b, 
 are also shown in Fig.~\ref{fig1}, in contrast to
 those of the

\begin{landscape}
\begin{table}[p]
\tbl{The
  inverse level density parameter $K$  (with errors $\Delta K$ 
  in parenthesis) in units of MeV,
   found by the  least mean-square 
   fit 
   for 
      $^{175-186}$Pt  in the low energy states ranges
      restricted by maximal values of the excitation energy
      having clear spins 
      (from ENSDF database\protect\cite{ENSDFdatabase}),  $U_{\rm max}$
          (also in MeV units),
       with  the
    precision of the standard expression for $\sigma$, Eq.~(\ref{chi}),
    are shown
    for 
  several approximations 
  with the same notations as in 
        Fig.~\ref{fig1}, 
       see also text.
       The MMA approaches are presented with 
       minimal $\sigma$ which 
     were  obtained
          for 
     the corresponding one-component 
     systems of $A$ nucleons\protect\cite{PRC}.
    \label{table-1}}
{
\begin{tabular}{@{}ccccccccccccccc@{}}\toprule
    & &
\multicolumn{2}{c}{FG}    &
\multicolumn{2}{c}{SFG}   &
\multicolumn{2}{c}{MMA1}  &
\multicolumn{2}{c}{MMA2a} &
\multicolumn{2}{c}{MMA2b} &
\multicolumn{3}{c}{One-component system} \\ \cline{3-12}
    &
\multirow{3}{*}{\minitab[c]{$U_{\rm max}$\\ MeV}} &
\multirow{3}{*}{\minitab[c]{$K$($\Delta K$)\\ MeV}} & &
\multirow{3}{*}{\minitab[c]{$K$($\Delta K$)\\ MeV}} & &
\multirow{3}{*}{\minitab[c]{$K$($\Delta K$)\\ MeV}} & &
\multirow{3}{*}{\minitab[c]{$K$($\Delta K$)\\ MeV}} & &
\multirow{3}{*}{\minitab[c]{$K$($\Delta K$)\\ MeV}} & & &
\multirow{3}{*}{\minitab[c]{$K$($\Delta K$)\\ MeV}} & \\
$A$ & & & $\sigma$ & & $\sigma$ & & $\sigma$ & & $\sigma$ & & $\sigma$ &
 Approach & & $\sigma$ \\[-1ex]
 &&&&&&&&&&&&&& \\ \colrule
175 & 1.74 &
13.2~(2.0) & 7.2 &  
13.2~(1.9) & 7.2 &  
12.0~(2.1) & 9.6 &  
19.2~(1.8) & 4.9 &  
31.0~(2.0) & 2.9 &  
MMA2b &
43.5~(2.4) & 2.3 \\
176 & 4.04 &
26.7~(1.7) & 5.6 &  
26.1~(1.6) & 5.5 &  
25.2~(1.8) & 6.5 &  
27.7~(1.3) & 4.8 &  
48.7~(1.4) & 2.3 &  
MMA2b &
65.6~(1.7) & 1.9 \\
177 & 0.43 &
~3.7~(0.5) & 4.3 &  
~3.7~(0.5) & 4.3 &  
~3.4~(0.5) & 4.9 &  
~6.3~(0.6) & 2.9 &  
15.7~(1.1) & 1.8 &  
MMA2b &
23.1~(1.6) & 1.7 \\
178 & 1.18 &
12.3~(1.0) & 2.1 &  
12.2~(1.0) & 2.1 &  
11.2~(1.0) & 2.7 &  
15.0~(0.8) & 1.6 &  
33.0~(1.2) & 0.8 &  
MMA2b &
47.2~(1.8) & 0.9 \\
179 & 0.73 &
~5.8~(0.6) & 4.9 &  
~5.8~(0.6) & 4.9 &  
~5.3~(0.6) & 5.8 &  
~8.1~(0.5) & 3.2 &  
18.7~(0.7) & 1.5 &  
MMA2b &
26.8~(1.0) & 1.4 \\
180 & 1.35 &
13.5~(0.7) & 1.8 &  
13.4~(0.7) & 1.8 &  
12.5~(0.8) & 2.4 &  
15.5~(0.6) & 1.5 &  
34.6~(1.1) & 1.0 &  
MMA2b &
49.2~(1.9) & 1.1 \\
181 & 0.32 &
~3.1~(0.2) & 2.2 &  
~3.1~(0.2) & 2.2 &  
~2.9~(0.2) & 2.7 &  
~5.1~(0.2) & 1.5 &  
13.0~(0.8) & 1.5 &  
MMA2b &
19.0~(1.3) & 1.6 \\
182 & 1.44 &
14.0~(0.6) & 1.6 &  
13.7~(0.5) & 1.5 &  
12.9~(0.7) & 2.2 &  
15.2~(0.4) & 1.3 &  
34.9~(1.4) & 1.4 &  
MMA2a &
18.5~(0.5) & 1.2 \\
183 & 0.59 &
~4.9~(0.5) & 5.8 &  
~4.9~(0.5) & 5.1 &  
~4.5~(0.5) & 6.2 &  
~6.6~(0.4) & 3.5 &  
17.1~(0.6) & 1.3 &  
MMA2b &
24.7~(0.8) & 1.2 \\
184 & 1.31 &
13.5~(0.6) & 1.4 &  
13.1~(0.6) & 1.6 &  
12.5~(0.7) & 1.9 &  
14.1~(0.5) & 1.4 &  
35.1~(2.1) & 1.8 &  
MMA2a &
17.1~(0.6) & 1.3 \\
185 & 0.73 &
~5.9~(0.5) & 4.4 &  
~5.8~(0.4) & 4.4 &  
~5.5~(0.5) & 5.3 &  
~7.2~(0.4) & 3.3 &  
18.2~(0.5) & 1.3 &  
MMA2b &
25.7~(1.1) & 1.8 \\
186 & 1.60 &
15.3~(0.7) & 1.9 &  
14.8~(0.6) & 1.8 &  
14.2~(0.8) & 2.5 &  
15.1~(0.5) & 1.8 &  
37.3~(1.3) & 1.2 &  
MMA2b &
15.1~(0.5) & 1.8 \\
187 & 0.47 &
~3.7~(0.4) & 4.7 &  
~3.7~(0.4) & 4.7 &  
~3.4~(0.5) & 6.0 &  
~5.0~(0.4) & 3.4 &  
14.2~(0.7) & 1.6 &  
MMA2b &
20.6~(1.0) & 1.6 \\
188 & 2.46 &
19.8~(0.9) & 2.8 &  
18.6~(0.7) & 2.7 &  
18.6~(1.0) & 3.5 &  
18.3~(0.6) & 2.8 &  
41.3~(1.2) & 1.7 &  
MMA2b &
56.7~(2.2) & 2.0 \\
189 & 0.26 &
~2.8 (0.1) & 1.3 &  
~2.8~(0.1) & 1.3 &  
~2.5~(0.3) & 3.2 &  
~4.1~(0.2) & 1.3 &  
12.6~(0.3) & 0.4 &  
MMA2b &
18.6~(0.5) & 0.5 \\
190 & 1.83 &
15.7 (0.3) & 1.8 &  
14.6~(0.3) & 1.6 &  
15.1~(0.3) & 1.6 &  
14.2~(0.2) & 1.8 &  
32.4~(1.7) & 4.1 &  
MMA1 &
18.0 (0.4) & 1.8 \\
191 & 0.56 &
~5.0~(0.7) & 6.1 &  
15.8~(0.4) & 1.4 &  
15.6~(0.5) & 1.7 &  
~6.2~(0.6) & 4.8 &  
18.0~(1.2) & 2.5 &  
MMA2b &
30.4~(1.0) & 1.7 \\
192 & 1.79 &
16.5 (0.9) & 2.6 &  
15.5 (0.8) & 2.6 &  
15.4 (0.9) & 2.8 &  
15.3 (0.7) & 2.6 &  
37.6~(2.4) & 2.6 &  
MMA1 &
18.0~(0.4) & 1.8 \\
193 & 0.27 &
~2.9~(0.4) & 3.2 &  
~2.8~(0.4) & 3.2 &  
~2.5~(0.5) & 4.9 &  
~4.0~(0.5) & 2.8 &  
15.2~(1.0) & 1.3 &  
MMA2b &
21.2~(1.5) & 1.2 \\
194 & 1.51 &
14.3~(0.9) & 2.5 &  
13.5~(0.8) & 2.5 &  
13.4~(0.8) & 2.5 &  
13.4~(0.7) & 2.5 &  
33.2~(2.9) & 3.1 &  
MMA1 &
18.0~(0.4) & 1.8 \\
195 & 1.02 &
~7.6~(0.6) & 5.8 &  
~7.5~(0.6) & 5.7 &  
~7.1~(0.6) & 6.7 &  
~8.0~(0.5) & 5.0 &  
21.6~(0.7) & 2.0 &  
MMA2b &
30.4~(1.0) & 1.7 \\
196 & 2.09 &
15.7~(0.3) & 1.8 &  
14.6~(0.3) & 1.9 &  
15.1~(0.3) & 1.6 &  
14.2~(0.2) & 1.8 &  
32.4~(1.7) & 4.1 &  
MMA1~ &
18.1~(0.4) & 1.8 \\
197 & 0.77 &
~7.2~(0.9) & 5.0 &  
~7.1~(0.8) & 5.0 &  
~6.6~(0.9) & 6.0 &  
~7.7~(0.7) & 4.6 &  
22.9~(1.4) & 2.1 &  
MMA2b &
32.9~(1.9) & 1.9 \\
198 & 1.37 &
15.6~(0.9) & 1.5 &  
14.5~(0.7) & 1.5 &  
14.5~(0.8) & 1.7 &  
13.9~(0.6) & 1.5 &  
38.4~(2.5) & 1.6 &  
MMA2a &
16.8~(0.7) & 1.5 \\
199 & 1.24 &
10.9~(2.3) & 7.0 &  
10.4~(2.0) & 6.9 &  
~9.5~(2.1) & 8.6 &  
~9.9~(1.6) & 7.3 &  
30.4~(3.7) & 3.9 &  
MMA2a &
43.6~(5.1) & 3.5 \\
200 & 1.84 &
19.7~(0.8) & 1.3 &  
17.6~(0.6) & 1.3 &  
18.4~(0.8) & 1.6 &  
16.5~(0.6) & 1.5 &  
45.3~(2.2) & 1.4 &  
MMA2b &
63.5~(3.6) & 1.6 \\ \botrule
\end{tabular}
}
\end{table}
\end{landscape}

\noindent
MMA1 approach.
 The results of the
 SFG asymptotical full saddle-point 
 approach, 
Eq.~(\ref{SPMgennp}), and of a similar popular  FG approximation, 
    Eq.~(\ref{FG}), 
which
are both in good agreement with those of the
standard Bethe formula\cite{Be36} for 
one-component systems 
    (see Ref.~\refcite{PRC}), are presented 
in Table \ref{table-1}.
      For finite realistic values of
      $\mathcal{E}_{\rm sh}$,  the value of the inverse level density
      parameter $K$ of MMA2a  (Table~\ref{table-1}) and the
      corresponding level density
      (Fig.~\ref{fig1})
      are in between those of the MMA1 and MMA2b. 
      Sometimes, the results of the MMA2a approach 
       are significantly
      closer to those of the MMA1 
      one, than to those of the MMA2b approach, e.g., for nuclei
      as $^{196}$Pt.

In 
both panels
of 
Fig.~\ref{fig1}, one can see the divergence
of the 
SFG, Eq.~(\ref{SPMgennp}), 
    level density asymptote  in
the zero excitation energy limit $U\rightarrow 0$. 
    This is clearly seen also analytically, in particular  
in the FG limit,  Eq.~(\ref{FG}); 
see also the general asymptotic  expression
(\ref{rhoasgennp}). 
It is, obviously, in contrast to any MMAs'
combinatorics
expressions 
(\ref{den0gennp})
in this limit;
see 
Eq.~(\ref{denbesnp}).
The MMA1 results are close to those of the 
 FG and SFG approaches 
    for all considered nuclei
    (Table \ref{table-1}), 
    in particular,
for both 
$^{195}$Pt and $^{196}$Pt \ isotopes 
in Table \ref{table-1}. The reason is that
their differences are
essential only for extremely small excitation energies $U$ where
 the MMA1 approach is finite while 
other,  FG and SFG,
approaches 
are divergent. 
However, there 
are almost no experimental data for 
 excited states
in the range of their differences, at least in the nuclei under
consideration.

    The MMA2b results,
  Eq.~(\ref{rho3-2b}), 
  for
$^{195}$Pt (see Fig.~\ref{fig1}(a)) with $\sigma \sim 1$
   are 
significantly better in agreement with the experimental
data
as compared 
to the results of all other approaches
(for the same nucleus).
 For this nucleus, the MMA1 (Eq.~(\ref{rho2})),
FG (Eq.~(\ref{FG})),
    and 
    SFG (Eq.~(\ref{SPMgennp})) 
approximations 
    are characterized by
    much larger $\sigma$  
     (see
    Table \ref{table-1}).
    In contrast to the case of 
    $^{195}$Pt (Fig.~\ref{fig1}(a)) with
    excitation energy spectrum 
     having a 
     large number of low energy states 
     below
    about 1 MeV, for
    $^{196}$Pt (Fig.~\ref{fig1}(b))  with almost no such states  in the same
     energy range, one 
    finds the opposite case~--~
    a significantly 
     larger MMA2b value of $\sigma$ 
     as compared to those
     for
    other approximations 
    (Fig.~\ref{fig2} and Table \ref{table-1}). 
    In
    particular, for  MMA1 (case i), and
    other asymptotic approaches
     FG and SFG, 
   one obtains  for $^{196}$Pt spectrum almost the same 
$\sigma \sim 1$,
and almost the same for 
MMA2a (case ii) with realistic values of
$\mathcal{E}_{\rm sh}$. 
 Again, notice 
that 
the MMA2a results (Eq.~(\ref{rho3}))
 are more close, at the realistic $\mathcal{E}_{\rm sh}$,
to  those of the MMA1 
(case i), as well as
 the results of the  FG and SFG approaches.
The MMA1 and MMA2a results (at
realistic values of $\mathcal{E}_{\rm sh}$) as well as 
 those of the
 FG and SFG approaches 
are
obviously much better in agreement with the
experimental data\cite{ENSDFdatabase} (see \ref{appC})
for $^{196}$Pt (Fig.~\ref{fig1}(b)).

One of the reasons of the exclusive properties of $^{195}$Pt 
 (Fig.~\ref{fig1}(a)),
    as compared to $^{196}$Pt (Fig.~\ref{fig1}(b)),
        might be assumed to be 
        the nature
    of the excitation energy
    in these nuclei.
    Our MMAs results (case i) or (case ii) could clarify the excitation nature
        as 
    assumed in Ref.~\refcite{PRC}.  
   Since 
    the  MMA2b results (case ii)  are 
    much better
    in agreement with the experimental data than the MMA1 results
    (case i)
    for $^{195}$Pt,
    one could presumably 
    conclude
    that for $^{195}$Pt 
    one finds more
    clear thermal low-energy excitations. 
    In contrast to this, for   $^{196}$Pt
    (Fig.~\ref{fig1}(b)), one observes more regular high-energy
    excitations 
    coming, e.g.,  
   from the dominating rotational energy
            $E_{\rm rot}$, 
   see Refs.~\refcite{KM79,PRC}.
   As seen, in particular, from the values of the inverse level
   density parameter $K$ and the shell structure of
   the critical quantity,
   $\xi^\ast \propto \mathcal{E}^{\prime\prime}_{\rm sh}/a =
    K \mathcal{E}^{\prime\prime}_{\rm sh}/A$,
Eq.~(\ref{pars}),
these properties can be 
understood 
to be mainly due to the larger values of
$K$ and shell correction second derivative $\mathcal{E}^{\prime\prime}_{\rm sh}$,
for low energy states 
in  $^{195}$Pt
 (Table~\ref{table-1})
 versus those of the $^{196}$Pt spectrum. 
 This is  in addition to the shell effects, which 
    are very important 
    for the case (ii) 
     which is not even realized 
    without    
  their dominance.

  As results, the statistically averaged level densities $\rho(E,N,Z)$
  for the MMA with a minimal value of the control-error
  parameter $\sigma$, Eq.~(\ref{chi}), in plots of
 Fig.~\ref{fig1} 
 agree well with those of the experimental data.
 The results of the MMA,  SFG and FG approaches
 for the level densities
 $\rho(E,N,Z)$ 
 in
Fig.~\ref{fig1}, and for $K$ in
Table~\ref{table-1},
do not depend 
on the cut-off spin factor
and moment of inertia 
because of the summations (integrations) over all spins, 
indeed, with accounting for the degeneracy $2I+1$ factor.
We do not use 
empiric free fitting parameters in our calculations, in particular, for
the FG results shown  in 
Table~\ref{table-1}, 
in contrast
to the back-shifted Fermi gas\cite{DS73}
and constant temperature models, see also
    Ref.~\refcite{EB09}.

The results of calculations 
for the inverse
level density parameter
$K$ 
in the long Pt isotope chain with $A=175-200$ are summarized in
Fig.~\ref{fig2}  and Table~\ref{table-1}. 
 Preliminary spectra data for nuclei far away from the
    $\beta$-stability line
     from
     Ref.~\refcite{ENSDFdatabase} are included in comparison with
     the results of the 
     theoretical approximations. These experimental data,
     may be incomplete.
    Nevertheless, 
    it might be helpful to present 
    the comparison between theory and experiment to check
    general common effects of the 
    statistical isotopic asymmetry and shell structure
    in a more wide
    range of nuclei around
    the $\beta$-stability line.
    
    As seen in Fig.~\ref{fig2},
        the results for $K$ for the isotopes of Pt
    ($Z=78$) as a function of 
    the particle number $A$  
    are characterized by a very pronounced saw-toothed behavior
with the alternating
low and high $K$ values for odd and even nuclei, respectively.
This behavior is more pronounced
for the MMA2b (close black dots) with larger $K$ values.
 For each nucleus, the significantly
smaller MMA1 value of $K$ (full red squares) 
is 
close to  that of 
the SFG (open green circles).
The FG results are 
    very close to those of the SFG approach and,
therefore, 
are not shown
in the plots, but presented in the Table. 
The MMA2a results for $K$ are intermediate between the MMA2b and MMA1 ones,
 but 
closer to the
MMA1 values.

 Notice that    for rather a long chain of the isotopes of Pt,
    one finds the remarkable shell oscillation
    (Fig.~\ref{fig2}).
Fixing the even-even (even-odd) chain, for all compared approximations,
one can see a hint of slow oscillations by evaluating 
its period  $\Delta A$,
$\Delta A \sim 
 40$ for $A\sim 200$
(see Ref.~\refcite{NPA} and Eq.~(\ref{dA})).
  Within order of magnitude,  these estimates agree, 
 with the
 main period for the relative shell corrections,
 $\delta E A/\tilde{E}$, 
shown in Fig.~\ref{fig3}(b). 
Therefore,
according to these evaluations and Fig.~\ref{fig3}(b)
for $K(A)$,  we show the sub-shell effects within a major shell. 
This shell oscillation as function of $A$ is
more pronounced for the MMA2b case because of its relatively  large amplitude,
but is 
mainly proportional to that of the MMA2a and other approximations.

The MMAs results shown in Fig.~\ref{fig2}, as function of
    the
    particle number $A$, Eq.~(\ref{pars}), 
 can be partially understood  through
 the 
 basic 
 critical quantity, $\xi\sim \xi^\ast \leq  \xi^\ast_{\rm max}$, where
     $ \xi^\ast_{\rm max} 
    \propto
       KU_{\rm max}\mathcal{E}^{\prime\prime}_{\rm sh}/A^{4/3}$~. 
       Here we need also  the maximal
      excitation energies
      $U_{\rm max}$ of the low energy states
      (from Ref.~\refcite{ENSDFdatabase} and Table \ref{table-1})
      used in our calculations. Such
      low energy states  
      spectra are more 
    complete due to information on
        the spins 
   of the states. 
   We extended the particle number interval beyond the range of
   known spectra
   in Fig.~\ref{fig3}(b) to 
    demonstrate a
  clear major shell.  
     As 
        assumed in the derivations
            (Subsection~\ref{subsec-MMA}), larger values of
    $\xi^\ast$, Eq.~(\ref{pars}), are 
   expected in the MMA2b
   approximation (see Fig.~\ref{fig2}), first of all because of 
   large $K$ (small level density parameter $a$).
   For the
        MMA1 approach,
     one finds significantly smaller 
         $\xi^\ast$, and 
         in between values 
     (more close to the MMA1)
     for the MMA2a case.
     This is  in line 
    with the assumptions for case (i) and case (ii)
     in the derivations of the
    MMA1, Eq.~(\ref{rho2}),
    and MMA2, Eq.~(\ref{rho3}),
    level-density approximations, respectively.

    In order to clarify 
        the
    shell effects, we present in Fig.~\ref{fig3}(a)
    the
    inverse level density parameters $K(A)$ 
     taking the MMA
        results with
        the smallest values of $\sigma$ at each nucleus 
        (see Fig.~\ref{fig2}
    and Table
    \ref{table-1}). 
     Among all MMAs results, this provides  
     the best agreement 
     with
     the experimental data for the statistical level density obtained
     by the sample method
    (\ref{appC}).
        The relative energy shell corrections\cite{MSIS12},
    $\delta E A/\tilde{E}\approx \delta E A/BE$, are presented
    by crosses
    in
    Figure \ref{fig3}(b).

    The oscillations in Fig.~\ref{fig3}(a) 
     are associated
    with sub-shell
        effects within
     the major
     shell, shown in Fig.~\ref{fig3}(b). As seen from
      Fig.~\ref{fig3} and Table \ref{table-1},
      the results of the
      MMA2b approach the better
          agree with experimental data 
      the larger
      number of states in the low energy states 
      range and the smaller maximal excitation
     energies, $U_{\rm max}$.  This is not the case 
     for the MMA1 and other approaches. One of the most pronounced
      cases
     was considered above for the $^{195}$Pt and $^{196}$Pt nuclei.
     However, 
     in the middle of the Pt chain,
     the results of
     all approximations are not well distinguished because of almost the same
     $\sigma$. 
     Except for the nuclei $^{176,188}$Pt,
     where one has relatively large
     $U_{\rm max}$, 
         for significantly smaller $U_{\rm max}$ the MMA2b approach is obviously
         better than
         other approaches.
    Most of the Pt isotopes under the consideration are well deformed,
     and the excited
     energy spectra (Ref.~\refcite{ENSDFdatabase})
     begin with relatively small energy levels in the low energy states 
     region.
     Therefore, 
      for simplicity, the pairing
     effects 
     were not taken into account in our Pt calculations, even in the simplest
     version\cite{PRC} of a shift of the excitation energy 
      by pairing condensation energy.
     We should emphasize
     once more that the MMA2 approach 
         for each of nucleus is important in the case (ii)
of dominating shell effects (see  Subsection~\ref{subsec-MMA}).
 As shown in the Table, 
 the isotopic 
  asymmetry 
 effects are important as compared to those of
the corresponding 
one-component nucleon 
case\cite{PRC}. They decrease
significantly
the inverse level density parameter $K$, especially for the MMA2b approach.

Thus, for the Pt isotope chain, one can clearly see  
                almost major shell region and
                   approximately constant 
               mean values of $K(A)$ for
               each approximation
                (Fig.~\ref{fig2}).        
  The MMA1 approach yields essentially small values 
  for $K$, which are 
  closer
  to that of the 
   neutron resonances. Their values
are 
a little smaller than those of the MMA2a,
and much smaller than
those of the MMA2b approach (Fig.~\ref{fig2}).
As seen clearly from  Figs.~\ref{fig2} and \ref{fig3}, and
    Table \ref{table-1}, 
    in line with 
    results of
 Ref.~\refcite{ZS16}, 
 the obtained values for $K$ 
 within the MMA2 approach can be  
essentially different from those of the MMA1 approach and those 
 of the SFG and FG approaches found, mainly, for the neutron resonances.
 Notice that, as in Ref.~\refcite{NPA}, in all our calculations of
    the statistical level density,
$\rho(E,N,Z)$, we did not use a popular assumption of small spins at large
excitation energies $U$, which is valid for the neutron resonances.
Largely speaking, for the MMA1 approach, one finds values for
$K$ of the same order as
those of the 
 FG and SFG approaches.
These mean values of $K$ are mostly 
close to 
 those of neutron resonances 
in order  of 
magnitude. For the 
FG and SFG approaches, Eqs.~(\ref{FG}) and
(\ref{SPMgennp}), respectively,
 can be understood because neutron
 resonances appear 
 relatively at large
excitation energies $U$. 
For  these  
resonances, as for MMA1,
    we should not expect 
    such strong shell effects as assumed to be in 
    the MMA2b approach. 
More systematic study of large deformations,  neutron-proton asymmetry, 
and 
pairing correlations 
 (Refs.~\refcite{Er60,Ig83,So90,AB00,AB03,ZH19,KZ20})
should be 
      taken into account to improve the comparison with experimental data,
      see also preliminary estimates in
 Ref.~\refcite{PRC} for 
  the rare earth and
      actinide nuclei.

      \section{Conclusions}

We derived the statistical level density $\rho(S)$ as function of the
entropy $S$
within the micro-macroscopic
approximation (MMA) using the mixed micro- and grand-canonical ensembles,
accounting for the neutron-proton asymmetry beyond
the standard saddle point method 
of the Fermi gas (FG) model.
This function can be applied for small and, relatively, large entropies
$S$, or excitation energies
$U$ of a nucleus. For a large entropy (excitation energy), one obtains
the 
exponential asymptote of the standard saddle-point
Fermi-gas model, however, with 
the significant inverse, $1/S$, power corrections.
For small $S$ one finds the usual finite combinatorics
expansion in  powers of $S^2$. Functionally, the MMA 
linear approximation in the $S^2 \propto U$
    expansion, at small excitation energies $U$,
    coincides with that of the empiric
    constant ``temperature'' model, 
    obtained without using 
    free fitting parameters. Thus,
    the MMA  unifies the 
    well-known Fermi-gas approximation with the 
    constant ``temperature'' model 
    for large and small entropies $S$, respectively,
    also with accounting for the neutron-proton asymmetry. 
The MMA at low excitation energies
    clearly manifests an 
advantage
over the standard full saddle-point 
approaches
because of 
no divergences of the MMA in the limit of
small
excitation energies, in contrast to all of full saddle-point method,
e.g., 
the Fermi gas 
 asymptote.
Another advantage takes place 
for nuclei which have
a lot of states in the very low-energy states 
range. In this case, the MMA 
results with only one physical parameter in the least mean-square fit, 
the inverse level density parameter $K$, 
is usually the better the larger number of the extremely low energy states
range.  These results are
certainly much better
than those for the 
Fermi gas model.  
The values of the inverse
level density
parameter K 
are compared with those of experimental data for low energy states 
below
neutron resonances 
in 
nuclear spectra of 
several nuclei. 
The 
MMA values
of 
$K$
for low energy states 
can be significantly
different from those of the neutron resonances within
the Fermi gas 
model.

 We have found 
 a significant shell effects in the MMA level density for the
 nuclear low-energy states 
 range
within the semiclassical periodic-orbit theory. 
In particular, we generalized the known saddle-point method 
results for the level density in
terms of the full saddle-point method  
shell-structure Fermi gas  (SFG) approximation,
accounting
for the shell, along with the neutron-proton asymmetry
effects
using the periodic-orbit theory. 
Therefore, a reasonable description of the
low energy states 
experimental data for the
statistical averaged level density,
obtained by the sample method
within the MMA with the help of the semiclassical periodic-orbit theory,
was achieved.
We emphasize the
importance of the shell, and neutron-proton  
effects in these
calculations.   
We obtained 
values of the inverse level
density parameter
$K$ for low-energy states 
range which are essentially different
from those 
of 
 neutron resonances. Taking a long  Pt isotope chain as a typical example,
one finds a saw-toothed behavior of $K(A)$
as function of the particle number $A$ and its 
    remarkable shell oscillation.
We obtained values of $K$
that are significantly 
larger 
than those obtained
for 
neutron resonances,  due mainly to 
accounting for  
the shell effects.
 We show that the semiclassical periodic-orbit theory is helpful in the
low-energy states range
  for analytical description of the level density
   and energy shell corrections. They are taken into account
   in the linear approximation up to small corrections due to the
   {\it residual} interaction beyond
   the mean field and extended Thomas-Fermi approximation
   within the shell-correction method,
   see Refs.~\refcite{BD72,BK72}. The main part of the
    inter-particle interaction is described in terms of 
   the extended Thomas-Fermi counterparts of the
   statistically averaged nuclear potential, and in particular,
   of the
   level density parameter.

 Our approach can be applied to the statistical analysis of
the experimental
data on collective nuclear states, in particular, for the nearest-neighbor
spacing distribution calculations within the Wigner-Dyson theory of quantum
 chaos\cite{Ze96,Ze16,GK11}.
As the semiclassical periodic-orbit 
MMA is the better the larger particle number
    in a Fermi system, one can apply this method 
    also for study of the metallic clusters and quantum dots
    in terms of the statistical level density, and  of several problems in
    nuclear astrophysics.
As perspectives,
the 
collective rotational excitations at
large nuclear angular momenta and
deformations, 
as well as more consequently 
pairing correlations, all with a more systematic accounting
for
the neutron-proton asymmetry, will be 
taken into account in a future  work.
    In this way, we expect to improve the comparison of the
theoretical evaluations
with
experimental data on the level density parameter 
significantly for energy levels below the neutron resonances.

\section*{Acknowledgements}

The authors gratefully acknowledge 
D.\ Bucurescu, R.K.\ Bhaduri, M.\ Brack,
 A.N.\ Gorbachenko, and V.A.\ Plujko
for creative discussions.
 This work was supported in part by the  budget program
"Support for the development
of priority areas of scientific researches",  the project of the
Academy of Sciences of Ukraine (Code 6541230, No 0120U100434).
S.\ Shlomo is partially supported by the US
Department of Energy under Grant no. DE-FG03-93ER-40773.

\appendix

\section{Residual interactions and Landau theory for Fermi-liquids}
\l{appA}
         In the following we should clarify the definition of the residual
         interaction in our semiclassical approach, based on the Landau-Migdal
          theory\cite{La58,MI67},
              in contrast to that of the difference between the Hamiltonian
              with inter-particle interactions and mean field potential.
                       For this purpose, one may take 
          a simple example of the semiclassical
         quasiparticle Landau theory for the Fermi
         liquids\cite{La58,AK59,PN66,BP91,HP93}, see also the
         review article\cite{MG14}.
         The distribution function $\Phi({\bf r},{\bf p},t)$ in the phase space variables
          of the position,
         momentum, and time,
         ${\bf r},{\bf p}$,  and $t$, respectively, can be presented as a
         sum of the Thomas-Fermi, $\Phi_{\rm \tt{TF}}$, and time-dependent
         quasiparticle,
         $\Phi_{\rm d}({\bf r},{\bf p},t)$, components.

              Note that for a semiclassical description of dense
           Fermi-liquid systems, Landau
           suggested\cite{La58} to expand a many-body
           distribution function near the Fermi surface for small
           quasiparticle excitations. The quasiparticles are defined as
           excitations of
           the Fermi-liquid system as a whole.
           These small quasiparticle excitations look very similar to the
           independent particles, but with the effective mass which is
           different from
           the particle mass by taking into account the effective
           inter-particle interaction
           amplitude. However, this single-quasiparticle picture
            is not expected to work 
           far from
           the Fermi surface.
           Therefore, within the Landau theory\cite{La58,AK59} for
           infinite Fermi-liquids and
           and within 
           the Migdal theory\cite{MI67,HS82} for finite dense Fermi
           systems and then,
           in the Strutinsky
           shell correction method\cite{St67,BD72,BK72},
           we have to use a {\it re-normalisation}.
           This is carried out by
           replacing the
            averaged single-quasiparticle distribution functions 
           by those of the
           extended Thomas-Fermi counterparts\cite{BG85,BB03},
           which are related 
           approximately to  the
           macroscopic liquid
           drop energy. The extended Thomas-Fermi counterparts
           account well mainly for a local
           inter-particle interaction through the Strutinsky's
           statistical averaging procedure, see
           Refs.~\refcite{BD72,St67}. 
         
          The component
         $\Phi_{\rm d}({\bf r},{\bf p},t)$ is
         a small dynamical correction which is the solution to the
         Landau-Vlasov equation, $d \Phi_{\rm d}/dt=I_{\rm d}$,
         with the integral collision term
         $I_{\rm d}$.
          The collision term $I_{\rm d}$
         in the $t_{\rm c}$ relaxation-time approximation\cite{HP93,MG14} takes
         into account  a small contribution of collisions of the
         {\it quasiparticles} as compared to other self-consistent
         Vlasov terms, $d\Phi_{\rm d}/dt \propto \omega \Phi_{\rm d}$,
         where $\omega$
             is the frequency of the harmonic oscillating motion.
             This is provided 
            by the rare-collision condition, typical for quasiparticle
             excitations in  nuclear matter, $\omega t_{\rm c} \gg 1$,
         that means a large relaxation time
         $t_{\rm c}$ with respect to the
         characteristic time $\omega^{-1}$ of the collective motion with a
        frequency $\omega$. Thus, the
          {\it residual} interaction in this approach is presented
          by 
             relatively small integral quasiparticle-collision term,
         $I_{\rm d} \sim \Phi_{\rm d}/t_{\rm c}$,  with respect to Vlasov terms,
             as 
         $1/(\omega t_{\rm c}) \ll 1$.
          In our approach,
         we neglect this small residual
         interaction as compared to the 
         contribution of the statistically averaged extended Thomas-Fermi
         component for the main
         part of the inter-particle interaction. 
          For simplicity, as the first step in our approach,
         a collective dynamical-quasiparticle component 
         of the
         distribution function, $\Phi_{\rm d}({\bf r},{\bf p},t)$,
         (e.g., collective rotations and vibrations)
         is also neglected here. Instead, 
         we account for the quasiparticle shell-correction
         contribution to the level density by using the
         Strutinsky shell correction method\cite{St67,BD72}, basically
         within the Landau-Migdal theory\cite{La58,MI67}. 

\section{
  Semiclassical periodic-orbit theory 
  for isotopic asymmetric system}
\l{appB}

The level density shell corrections for neutron and proton systems can be presented analytically
within the periodic-orbit theory 
in terms of  the sum over classical periodic orbits\cite{SM76,BB03,SM77,MY11},
\be\l{goscsemnp}
\delta g^{(\tau)}_{\rm scl}(\varepsilon)=
\sum^{}_{\rm PO}g^{(\tau)}_{\rm PO}(\varepsilon),\quad
g^{(\tau)}_{\rm PO}(\varepsilon)=\mathcal{A}^{(\tau)}_{\rm PO}(\varepsilon)
~\cos\left[\frac{1}{\hbar}\mathcal{S}^{(\tau)}_{\rm PO}(\varepsilon)-
\frac{\pi}{2} \mu^{(\tau)}_{\rm PO}
-\phi^{(\tau)}_0\right].
\ee
Here $\mathcal{S}^{(\tau)}_{\rm PO}(\varepsilon)$ is the classical action along the
periodic orbit
in the neutron ($\tau=n$) or proton  ($\tau=p$) potential well of the
same radius,
$R=r_0A^{1/3}$ ($r^{}_0\approx 1.1$ fm),
$\mu^{(\tau)}_{\rm PO}$ is the so called Maslov index, determined by
the catastrophe points (turning and caustic points) along the periodic orbit,
and
$\phi^{(\tau)}_0$ is an additional shift of the phase coming from the dimension
of the problem and degeneracy of the periodic orbits. 
The amplitude
$\mathcal{A}^{(\tau)}_{\rm PO}(\varepsilon)$, and the
action $\mathcal{S}^{(\tau)}_{\rm PO}(\varepsilon)$,
 are 
smooth functions of
the energy
$\varepsilon$. 
In addition, the amplitude,
    $\mathcal{A}^{(\tau)}_{\rm PO}(\varepsilon)$,  depends on
the 
periodic-orbit stability factors.
      The Gaussian local averaging of the level density shell
       correction, 
          $\delta g^{(\tau)}_{\rm scl}(\varepsilon)$, over the
        single-particle energy spectrum $\varepsilon^{}_i$ near the Fermi
       surface $\varepsilon^{(\tau)}_F$,
      with a width parameter
      $\Gamma$,
          smaller than a distance between major shells,
    $\mathcal{D}_{\rm sh}$, 
            can be done
  analytically\cite{SM76,BB03,MY11},
\be\l{avdennp}
\delta g^{(\tau)}_{\Gamma\;\rm scl}(\varepsilon) \cong
\sum^{}_{\rm PO}g^{(\tau)}_{\rm PO}(\varepsilon)~
\exp\left[-\left(\frac{\Gamma t^{(\tau)}_{\rm PO}}{2\hbar}\right)^2\right]~,
\ee
where $t^{(\tau)}_{\rm PO}=\partial S^{(\tau)}_{\rm PO}/\partial \varepsilon$
is the period
of particle motion along the periodic orbit 
in the corresponding potential well.

The 
smooth 
ground-state 
energy of the nucleus is 
approximated by
 $\tilde{E}_\tau\approx E^{(\tau)}_{\rm \tt{ETF}}=\int_0^{\lambda_\tau}
  \d \varepsilon~\varepsilon~ \tilde{g}_\tau(\varepsilon)$~,
   where
  $\tilde{g}_\tau(\varepsilon)$
   is a smooth 
  level density equal approximately to the extended Thomas-Fermi 
  level density, $\tilde{g}_\tau\approx g^{(\tau)}_{\rm \tt{ETF}}$,
   ($\lambda \approx \tilde{\lambda}$, 
  and $\tilde{\lambda}$ is the
  smooth  chemical potential  in the
  shell corrections method). 
 The chemical potentials $\lambda_\tau$ (or  $\tilde{\lambda}_\tau$) are
 the solutions of the corresponding conservation of particle number
  equations:
\be\l{chempoteqnp}  
   N = \int\limits_0^{\lambda_n}\mbox{d} \varepsilon~g_n(\varepsilon) ~,\qquad
  Z = \int\limits_0^{\lambda_p}
  \mbox{d} \varepsilon~g_p(\varepsilon)~.
  \ee

  The periodic-orbit 
  shell component of the free energy,
$\delta F^{(\tau)}_{\rm scl}$,  Eq.~(\ref{FESCFnp}),
is related in the non-thermal and non-rotational limit to the 
     shell correction energy of a cold nucleus,
    $\delta E^{(\tau)}_{\rm scl}$, see Eq.~(\ref{dEPO0Fnp}) and
    Refs.~\refcite{SM76,BB03,MY11}.
    Within the 
    periodic-orbit theory, $\delta E^{(\tau)}_{\rm scl}$
is determined, in turn, through Eq.~(\ref{dEPO0Fnp}) by 
the oscillating level density 
$\delta g^{(\tau)}_{\rm scl}(\varepsilon)$, 
see Eq.\ (\ref{goscsemnp}).

The chemical potential $\lambda_\tau $ can be approximated by the Fermi energy
$\vareps^{}_F$, up to small excitation-energy, 
and isotopic asymmetry
corrections ($T\ll \lambda_{\tau}\approx \lambda$
for the saddle point 
value $T=1/\beta^\ast$, if exists). 
It is determined by the particle-number conservation conditions, Eq.~
(\ref{chempoteqnp}),
  where
  $g_\tau(\vareps)\cong g^{(\tau)}_{\rm scl}=
  g^{(\tau)}_{\rm \tt{ETF}} +\delta g^{(\tau)}_{\rm scl}$
  is the total semiclassical  
  level
density of the periodic-orbit theory.
One now needs to solve equations (\ref{chempoteqnp}) to
determine the 
chemical potentials $\lambda_\tau $ as
  functions
  of the neutron $N$ and proton $Z$ numbers, 
  since 
$\lambda_\tau$  are needed in Eq.\  (\ref{dEPO0Fnp}) to 
obtain the semiclassical energy shell correction
$\delta E^{(\tau)}_{\rm scl}$. Neglecting a difference between the
$\lambda_\tau$
 at small asymmetry parameter $X^2$ as
  functions
  of the 
  particle numbers $A$,
one has to solve only one conservation equation for a mean $\lambda$,
 $~ A = \int\limits_0^{\lambda}\mbox{d} \varepsilon~g(\varepsilon)$~.

For a major shell structure near the Fermi energy surface
$\varepsilon\approx \lambda$,
the 
periodic-orbit energy shell correction,  $\delta E^{(\tau)}_{\rm scl}$
(Eq.~ (\ref{dEPO0Fnp})),
is approximately
proportional to the level density shell correction, 
$\delta g^{(\tau)}_{\rm scl}(\varepsilon)$ 
(Eq.\ (\ref{goscsemnp})), at $\varepsilon=\lambda_\tau$. 
Indeed, the rapid convergence of the 
 periodic-orbit sum in Eq.~(\ref{dEPO0Fnp})
is guaranteed by the 
factor in front of the density component $g^{(\tau)}_{\rm PO}$,
Eq.\ (\ref{goscsemnp}), a factor 
which is inversely proportional to the period
time $t^{(\tau)}_{\rm PO}(\lambda_\tau)$ squared along 
the periodic orbit. 
Therefore, only periodic orbits 
with 
short periods which occupy a 
significant 
phase-space volume near the Fermi surface will contribute.
These orbits are responsible for the
major shell structure, that is related to a Gaussian averaging 
    width,
$\Gamma\approx \Gamma^{(\tau)}_{\rm sh}$, which is much larger
    than the distance between neighboring 
    single-particle states but much smaller
than the distance
$D^{(\tau)}_{\rm sh} $ between major shells near the Fermi surface.
Eq.~(\ref{avdennp}) for the averaged 
single-particle level density
    was derived 
    under these
conditions for $\Gamma$.
According to the periodic-orbit theory\cite{SM76,BB03,MY11},
the distance between major shells, $D^{(\tau)}_{\rm sh}$, is
determined by a
mean period of the shortest
and  most degenerate  periodic orbits, 
$\langle t^{(\tau)}_{\rm PO}\rangle$:
\cite{SM76,SM77}
\be\l{periodenp}
\mathcal{D}^{(\tau)}_{\rm sh} \cong 
\frac{2\pi \hbar}{\langle t^{(\tau)}_{\rm PO}\rangle} 
\approx \frac{\lambda_\tau}{A^{1/3}}~.
\ee
The period, $\Delta A$, of the oscillating part, $\delta K(A)$, of the
inverse level density
parameter,
$\delta K\propto -A\delta g/\tilde{g}^2 $
is approximately defined by the shell structure period,
$\mathcal{D}_{\rm sh}\approx \mathcal{D}^{(\tau)}_{\rm sh}$, of the 
single-particle level-density shell correction,
Eq.(\ref{periodenp}),
as function of $\lambda$ ($\lambda \approx \lambda_\tau $): 
    See Eq.~(\ref{denparnp}) for
    $a_\tau\approx a/2$
and Eq.~(\ref{goscsemnp}) for $\delta g(\lambda)$.
Determining 
the particle number variable $A$ from the value of $\lambda $, 
one obtains
\be\l{dA}
\Delta A \approx \mathcal{D}_{\rm sh}\tilde{g} \sim 
 A^{2/3}~,
\ee
where the Thomas-Fermi 
estimate,
$\tilde{g} \approx g^{}_{\rm TF}\sim A/\lambda$,
was used.

Taking the factor in front of
$g^{(\tau)}_{\rm PO}$, in Eq.~(\ref{dEPO0Fnp}), 
    off the sum over the POs for
the energy shell correction
 $\delta E_{\rm scl}$, 
one arrives at
    its semiclassical expression\cite{BB03,SM76,SM77,MY11}
(\ref{dedgnp}). 
Differentiating Eq.~(\ref{dEPO0Fnp}) with
    (\ref{goscsemnp}) with respect to $\lambda_\tau$ and
keeping only the dominating terms coming from 
differentiation of the cosine of the action phase argument,
$S^{(\tau)}/\hbar \sim A^{1/3}$, one finds the useful relationships: 
\be\l{d2Edl2}
\frac{\partial^2\delta E^{(\tau)}_{\rm PO}}{\partial\lambda_\tau^2}
\approx -\delta g^{(\tau)}_{\rm PO}~,\quad
\frac{\partial^2 g_\tau}{\partial\lambda_\tau^2}\approx
\sum^{}_{\rm PO}\frac{\partial^2\delta g^{(\tau)}_{\rm PO}}{\partial\lambda_\tau^2}
\approx -\left(\frac{2\pi}{D^{(\tau)}_{\rm sh}}\right)^2
\delta g_\tau(\lambda_\tau)~.
  \ee

\section{The sample method}
\l{appC}

The statistical level density, $\rho(E,N,Z)$, as function of the
excitation energy $U$
can be calculated\cite{So90,LLv5} directly from the experimental  data on 
the energy-spin states, $\{U_i, I_i\}$, as
    $\rho_i^{\rm exp}=N_i/U_s$, where $N_i$
is the number of states in the $i$-th sample, $i=1,2,...,\aleph$,
 and $U_s$ is the sample energy length.
    The black dots in Fig.~\ref{fig1} are plotted at mean positions
of the experimental excitation energies for each $i$-th sample.

Convergence of the sample method over the equivalent sample-length
parameter $U_s$ of the statistical averaging 
was conveniently studied through the 
sample number $\aleph$, for a given spectrum. We assume that
     the statistical {\it plateau}  condition on a
      small change of  the single parameter $K$
      obtained by the least mean square fit 
      under variations of $\aleph$ is valid.
      The statistical condition, $N_i\gg 1$ for all $i$, with 
      $\aleph\gg 1$  on the plateau,
determines the accuracy
    of our calculations. Under these conditions 
     all microscopic details 
    can be neglected.
 For a given spectrum, 
    under these
    statistical plateau conditions, 
   the number $\aleph$ (or the sample lengths $U_s$)
    plays the 
role which is similar to that of 
averaging parameters (Gaussian widths and correction polynomial degrees)
in the 
 Strutinsky's smoothing procedure 
for the 
calculations of the averaged 
single-particle level density
\cite{BD72}.
In our case, the plateau condition means almost constant values
of the critical physical parameter $K$ for different variations of
 $\aleph$. 
 Therefore, the results
of Table
\ref{table-1}, 
calculated at the same  values of the found plateau, 
do not depend practically, with the statistical accuracy, on the
averaging
parameter $\aleph$
within the plateau.  Good plateau condition was obtained in a wide range around the
values near $\aleph\approx 5-7$. This is 
 an analogue of 
the energy and the level density
shell corrections, in which both are  independent of the Strutinsky's
    smoothing parameters.

The standard least-mean-square fit 
determines the applicability 
of
the theoretical 
approximations for $\rho(U_i)$ (Subsection~\ref{subsec-MMA}) 
for 
describing, in terms of {\it one} parameter $K$, the experimental
data \cite{ENSDFdatabase}, $\rho_i^{\rm exp}$, obtained
by the sample method from the nuclear excitation spectra.
    We realize this with the help of the
    least-mean-square fit
    control relative-error 
parameters:
\be\l{chi}
\sigma^2=\frac{\chi^2}{\aleph-1}, \qquad \chi^2 =\sum_i
\frac{(y(U_i)-y^{\rm exp}_i)^2}{(\Delta y_i)^2},
  \ee
where 
$y=\ln\rho$~.  Notice that the fitting parameter, $K=A/a$,
has a clear physical meaning as
related to the 
single-particle level density
$g(\lambda)$, modified by the Strutinsky's shell correction method\cite{BD72},
through Eq.~(\ref{denparnp}). But still the nuclear mean-field parameters
 for this level density calculations can be
varied within the least-mean-square fit 
for the description in terms of the statistical level density.
For units of the theoretical versus experimental differences,
$y(U_i)-y^{\rm exp}_i$, one may evaluate
$\Delta y_i = \Delta \rho_i/\rho_i\sim 1/\sqrt{N_i}$ 
through  the  dispersion, $\Delta \rho_i^2 \sim N_i$ 
in the statistical distributions of
    the excitation states of the available experimental data over samples. 
Restrictions which connect spin projections $M_i$ of
the state spin $I_i$ with the spin itself, in our problem with the spin degeneracy $2I_i+1$
of the quantum states,
can only diminish $\Delta \rho_i$.
Therefore,
$\sqrt{N_i}$ 
 is the maximal statistical-error
 estimate for $\Delta \rho_i$
 as compared to 
those with these restrictions. Such
errors  are convenient to use
in Eq.~(\ref{chi})  as units of $y(U_i)-y^{\rm exp}_i$.
Notice that taking other units, e.g., $\Delta y_i=1$,
one obtains almost the same
    least-mean-square fit results for $K$
   and ratio of $\sigma$ for different approximations with accuracy much 
   better than 20\%.

    We  calculate $\sigma$, Eq.~(\ref{chi}), at the minimum of
    $\chi^2$ over the unique parameter, $K=K_{\rm min}$,
    having a definite physical meaning as the
    inverse level density parameter $K$.
     Then, we may compare these $\sigma$ values of
  several different MMA approximations, found independently on the data
  under certain conditions,
  with the well-known 
   Fermi gas approach. For this aim we are interested in
  relative values of 
  $\sigma(K_{\rm min})$ for different compared approximations
  rather than their absolute values. 
      Except for testing by Eq.~(\ref{chi}) in terms of $\sigma$,
  one should take into account that the theoretical approximations are valid
  for small 
    $\xi^\ast$ ($\xi^\ast \ll 1$, MMA1), or $1/\xi^\ast$ ($\xi^\ast \gg 1$, MMA2),
  see Subsection~\ref{subsec-MMA}.
   Therefore,  these conditions are also important, along with
      the found values of $\sigma$, to  
    determine
    a scatter of experimental points around a line of the theoretical
    approximation. 

%


\end{document}